\documentclass[10pt,aps,prd,longbibliography,reprint,onecolumn,notitlepage]{revtex4-1}
\usepackage{amsmath}
\usepackage{amssymb}
\usepackage{bm}
\usepackage{tensor}
\usepackage{graphicx}
\usepackage[bookmarks=false]{hyperref}

\providecommand{\ed}{\mathrm{d}}
\providecommand{\mA}{\mathbf{A}}
\providecommand{\mF}{\mathbf{F}}
\providecommand{\mG}{\mathbf{G}}
\providecommand{\mJ}{\mathbf{J}}

\providecommand{\mc}{\boldsymbol{\chi}}

\begin{document}

\title{Covariant kinematics of light in media and a generalized Raychaudhuri equation}
\author{Robert T. Thompson}
\email{robert@cosmos.phy.tufts.edu}
\affiliation{Institute of Applied Physics, 
Karlsruhe Institute of Technology (KIT),
76128 Karlsruhe, Germany} 

\author{Mohsen Fathi}
\email{m.fathi@shargh.tpnu.ac.ir}
\affiliation{Department of Physics,
Payame Noor University (PNU),
P.O. Box 19395-3697,
Tehran, Iran}

\date{\today}

\begin{abstract}
There is ongoing interest in adopting various tools and ideas from general relativity for optical applications and the study of light propagation through natural or engineered media.
Here, the covariant kinematics of light propagating through arbitrary dielectric media in curved space-times are derived, allowing for analysis and tracing of congruences of light through media that may smoothly vary in character between vacuum, positively refracting, and negatively refracting; or null, timelike, and spacelike with respect to the background metric.  
The kinematics are then used to generalize the Raychaudhuri equation -- an important tool in general relativity that describes the focus of a congruence. 
These results will be useful for the analysis of optical devices, particularly those designed using transformation optics, and serve as theoretical tools to study generalized concepts in general relativity.


\end{abstract}

\maketitle

\section{Introduction}
The last decade has seen a resurgence of interest in the covariant, tensorial description of electrodynamics in dielectric media.  
This has been largely spurred by developments in transformation optics \cite{Leonhardt2006njp1}, which uses longstanding ideas about the similarity of the refractive properties of dielectric media to those of curved space-times to design optical media with unusual features or capabilities, such as invisibility cloaks \cite{Pendry2006sc,Schurig2006sc}.
Other interesting applications of transformation optics, such as space-time event cloaking \cite{McCall2011jo}, highlight the potential of the general relativity-inspired formulation of transformation optics, and there is wide interest in what other ideas and tools from general relativity may be imported for optical applications.

Transformation optics as it stands is limited in its capacity to model and control real-world phenomena, such as the dispersion, dissipation, and nonlinearities that are manifest in the metamaterials used to actually realize transformation devices.
Progress incorporating these features in transformation optics has been made by formulating and manipulating tensorial versions of their descriptions \cite{Gratus2016njp,Paul2012oe,Bergamin2011prb}, which further encourages the development of covariant electrodynamics in media.
It is in the context of electrodynamics within dielectric media residing in a curved background space-time that we study one of the fundamental tools of general relativity: the kinematics of a congruence of light.

While the fantastic technological possibilities proffered by metamaterials and transformation optics are a significant motivator, there is also a great deal of overlap with other topics of theoretical interest.
Analog models of curved space-times seek to represent aspects of extreme gravitational systems -- e.g.\ light propagation near black holes -- with some kind of laboratory-accessible system \cite{Unruh1981prl,Unruh1995prd,Reznik2000prd,Schutzhold2002prl,Novello2002,Barcelo2005lrr,Thompson2010prd}.  
Mathematically, they are very naturally related to ideas in transformation optics and metamaterials \cite{Greenleaf2007prl,Smolyaninov2013pra,Chen2010oe,Lu2010jap,Boston2015prd,Narimanov2009apl}, which has resulted in a great deal of cross-fertilization of ideas.  
The tensorial, differential forms formulation of electrodynamics used here is also very closely related to developments in the premetric electrodynamics program, which seeks a deeper understanding of the structure and pliability of electrodynamics by considering the metric as a subsidiary field that does not directly enter into Maxwell's equations \cite{Hehl2002ijmpa,Hehl}. 
Lastly, a major focus of research in the area of quantum gravity and physics beyond the standard model is based on Lorentz violating space-times, and it has been shown that Lorentz violation in effective field theories is connected to Finsler geometries \cite{Kostelecky1989prd,Girelli2007prd,Kostelecky2011plb}.
The idea that space-time is Lorentzian was inspired by Maxwell's equations in vacuum \cite{Einstein1905adp}, but
one of the interesting features of refractive media is that, as discussed below, the light cone is determined by what can be considered an effective Finslerian structure -- the optical metric -- that provides a natural and accessible context for studies in Finsler-type space-times \cite{Perlick_Ray_Optics,Lammerzahl2004prd,Skakala2009jp}.

What ties these topics together in this work is the formal study of electrodynamics on manifolds that possess the additional structure of a dielectric medium.  
Since the vacuum can itself be considered a trivial dielectric medium, allowing for a nontrivial dielectric structure is a natural generalization.
The similarity between light in curved space-times and light in refractive media has been recognized since the early days of general relativity \cite{Eddington1920,Gordon1923}, and an identification between the two by Plebanski \cite{Plebanski1960pr} allowed the specification of a dielectric analog model of curved vacuum space-times \cite{DeFelice1971gerg} and was used as a foundation for transformation optics \cite{Leonhardt2006njp1}.
The point of departure made here compared with earlier related formulations or identifications of electrodynamics in media is an explicit recognition that any real medium resides in a space-time of which the curvature should be specifically accounted for, and a desire to disentangle the refractive action of the medium from that of the background space-time in which it resides.
For example, the well-known book by Post \cite{Post} considers a tensorial formulation of electrodynamics in media but does not include separate information about the background metric -- an implicit assumption of flat space-time -- while the premetric electrodynamics program uses a differential forms formulation of electrodynamics but considers a general structural field with no specified metric.
This could be interpreted as combining space-time and medium contributions, but for a rigorous formulation of transformation optics or study of light propagation in media it is desirable to distinguish the medium from the background space-time \cite{Thompson2011jo1,Thompson2011jo2,Thompson2015pra,Fathi2016prd}.

\section{Goal of this paper}

Consider a laser beam.  Physically, one might think of the laser beam as defining a column of light occupying some physical region of space.
The first question one might ask is what path the beam follows.  
The next question one might ask is how the shape of the beam changes along its path.  
To be more specific, one might want to measure the shape of the transverse cross section of the beam at one instant of time, nominally representing the shape of a phase front, and then compare this with the shape of a transverse cross section of the beam at a different position or later time.
Has the beam expanded, contracted, rotated, sheared from a circle to an ellipse?
We may think of the laser beam as made up of a bundle of curves, called a congruence, and we wish to characterize the transverse behavior, the kinematics, of this congruence.
Thus we wish to go beyond the idea of ray trajectories to study these other physically meaningful and measurable characteristics of a beam.

The kinematics of congruences is a fundamental tool in general relativity.  Not only is it important for accurate ray tracing, as displayed by the spectacular rendering of a black hole in the film \textit{Interstellar} \cite{James2015cqg}, but it is also fundamental in determining whether, and what type of, a horizon exists \cite{HawkingEllis}.
The associated Raychaudhuri equation gives information on the evolution of the cross sectional area of the congruence -- essentially how the focus changes along the length of the beam -- and is the basis of the focusing and singularity theorems \cite{HawkingEllis}.

There are differences between null, timelike, and spacelike congruences (e.g.\ in the dimensionality of their transverse subspaces), and while the strategy for studying their kinematics is the same in each case, they must be done separately.  
On the other hand, light propagating inside dielectric media typically moves slower (or sometimes faster, which is not strictly forbidden by causality for heavily dispersive media) than the vacuum speed of light.
Given that metamaterials allow the possibility (at least in principle) for light rays to smoothly transition between null, timelike, and even spacelike propagation with respect to the background metric, it is natural to ask whether one must shift between the null, timelike, and spacelike pictures of kinematics as the congruence is traced, or whether there exists some unified approach that allows us to continuously follow a congruence of light through media.

The typical textbook derivation (see, e.g., Ref.~\cite{Poisson}) of the kinematical decomposition of a null congruence assumes that it is both geodesic and null with respect to the space-time metric.
Since neither of these conditions hold within refracting media the textbook derivation is not valid for our purpose. 
To address this more general question of non-null, non-geodesic light propagation in media we develop the theory \textit{ab initio} from Maxwell's equations with two main goals in mind.

First, we seek to derive the kinematics of a congruence of light passing through linear dielectric media residing in a potentially curved background space-time. 
This will allow a more complete analysis and congruence tracing through media, which should be useful for fields like transformation optics.

Second, we find a generalized form of the Raychaudhuri equation that is valid for congruences in linear dielectric media residing in a potentially curved background space-time.
For affinely parametrized null geodesics in curved \textit{vacuum} space-times, the Raychaudhuri equation is
\begin{equation}
 \dot{\Theta} = -\frac12 \Theta^2 - \hat{B}\indices{^{\alpha}_{\beta}} \hat{B}\indices{^{\beta}_{\alpha}} - R\indices{^{\rho}_{\lambda\rho\tau}}u^{\lambda}u^{\tau}.
\end{equation}
This equation provides information about the change in the cross sectional area $\Theta$ of a beam as one follows the beam; i.e., it describes the focus of the beam along its length.
Aside from interest in focusing from an optics point of view, a generalized Raychaudhuri equation is important to the theory of light in media because it contains the Riemann tensor and is therefore a measurable probe of curvature.
We expect that, given an arbitrary refractive medium, additional medium-dependent terms should appear in this generalized Raychaudhuri equation.
Given the longstanding idea that linear dielectric media is equivalent to a vacuum space-time, one might conjecture that these medium contributions will appear in the form of a tensor akin to the Riemann tensor, comprised of derivatives of the optical metric,  which does not vanish even in flat space-time.
Although we indeed find medium-dependent contributions to the Raychaudhuri equation, we do not find that they appear in the form of a Riemann-like tensor.

It was mentioned above that one of the interesting features of light in media is that the light cone is determined by an effective Riemann-Finsler ``optical metric.''
Recently there has been interest in the Raychaudhuri equation in Finsler space-times \cite{Stavrinos2005ijtp,Stavrinos2012gerg,Minguzzi2015cqg}, but our result turns out to be quite different.
The reason is that, while the lightcone is determined by the effective pseudo-Finslerian optical metric, this is not the only structure present in the system, and a real-world observer actually makes measurements with respect to the background metric.
Thus, while the optical metric is interesting in that it provides a structure that can be interpreted as Finslerian, the medium does not fundamentally behave like a vacuum pseudo-Finsler manifold since the underlying space-time structure is still Riemannian.

The paper is organized as follows. In Sec.\ \ref{Sec:ClassicalEM} we review the covariant formalism of light in media that we work with and take the ray-optics limit.  
In Sec.\ \ref{Sec:Kinematics} we derive the first order kinematics of a light congruence in media.  
In Sec.\ \ref{Sec:Raychaudhuri} we derive a generalized Raychaudhuri equation for light in media within curved space-times, which reduces to the usual Raychaudhuri equation in vacuum.
The main finding in this section is that the medium does not contribute a term that might be interpreted as an effective curvature tensor -- even for impedance matched media where the optical metric becomes purely pseudo-Riemannian.
We conclude with Sec.\ \ref{Sec:Conclusions}.

\section{Geometrical optics limit of covariant electrodynamics in media} \label{Sec:ClassicalEM}

Using the geometric optics limit of classical electrodynamics, we seek to study the kinematical behavior of a congruence of light rays traversing an arbitrary dielectric medium residing in a possibly curved background space-time.
Furthermore, we are interested in working in a totally covariant, four-dimensional tensorial framework based on differential forms.
In this section we review such a framework for electrodynamics in media and its geometric optics limit.
A more complete introduction to tensorial electrodynamics may be found in, e.g., Refs.~\cite{Post,MTW,Baez,Thompson2012aiep}.
To be clear, although in the Introduction we mentioned transformation optics and dielectric analog space-times as motivations for studying covariant electrodynamics in media on manifolds, we make no assumptions on the form of the medium under consideration; we simply want to study the propagation of light through an arbitrary slab of stuff residing in a curved background space-time.  

\subsection{Covariant electrodynamics in media}

Space-time curvature affects the propagation of light. 
The deflection of starlight passing near the Sun provided the first observational confirmations of general relativity, and such gravitational lensing is commonly exploited in modern astronomical and cosmological studies.
In Maxwell's equations, information about the space-time is entirely contained in an operator $\star$, called the Hodge dual.
All pseudo-Riemannian manifolds endowed with a metric, e.g., space-time, have a naturally associated Hodge dual constructed from the metric, and thus the Hodge dual contains all the space-time information of the metric tensor.
For $k<m$, the Hodge dual takes a $k$-form to an $(m-k)$-form;
in the case of a four-dimensional space-time, it maps a 2-form into another 2-form and has the index expression
\begin{equation}\label{Eq:HodgeCompon}
\star\indices{_{\alpha\beta}^{\mu\nu}}=\frac{1}{2}\sqrt{|g|}\,\,\epsilon_{\alpha\beta\sigma\rho}g^{\sigma\mu}g^{\rho\nu},
\end{equation}
where $g^{\mu\nu}$ is the contravariant form of the metric and $\epsilon_{\alpha\beta\sigma\rho}$ is the completely antisymmetric Levi-Civit\'{a} symbol.
The electromagnetic fields $(\vec{E},\vec{B})$ are components of an antisymmetric second rank covariant tensor (2-form), $\mF = F_{\mu\nu}$, and
Maxwell's homogeneous and inhomogeneous equations are \cite{MTW,Baez}
 \begin{align}
  \ed \mF &= 0, \\
  \ed \star \mF & =\mJ,
 \end{align}
where $\ed$ denotes the exterior derivative and $\mJ$ is the charge-current 3-form.
In the presence of polarizable media, an incident field $\mF$ can induce multipole moments that contribute to $\mJ$ even though the total monopole or free charge contribution to $\mJ$ may be zero.
The solutions to Maxwell's equations then include the particular solution
\begin{equation}
 \ed \star \mathbf{P} = \mJ_{multipole},
\end{equation}
in which case the general solution satisfies
 \begin{align}
  \ed \mF &= 0, \\
  \ed \star (\mF+\mathbf{P}) & = \ed\mG  =\mJ_{free}.
 \end{align}
We typically work with macroscopically neutral media where $\mJ_{free}=0$, but this is not necessary.
In media where the polarization response is linear in the applied field, the particular solution $\mathbf{P}$ is linear in the homogeneous solution $\mF$, and we may write \cite{Thompson2011jo1}
\begin{equation} \label{Eq:constitutive}
  \mG = \star\mc\mF
\end{equation}
where $\mc$ contains all the information about the medium such as permeability and permittivity in a metric-independent way.

As an alternative and useful interpretation, one could assume the fields $\mF$ and $\mG$ are independent and contain the components of $(\vec{E},\vec{B})$ and $(\vec{D},\vec{H})$, respectively.  Then solutions to Maxwell's equations $\ed\mF=0$ and $\ed\mG=\mJ_{free}$ require a constitutive relation that is $\mG=\star\mF$ in vacuum but which can be uniquely extended to Eq.~(\ref{Eq:constitutive}) in the presence of linear media.
In component form, the constitutive relation reads
\begin{equation}\label{Eq:G=StarFCompon}
   G_{\mu\nu}=\star\indices{_{\mu\nu}^{\alpha\beta}}\chi\indices{_{\alpha\beta}^{\sigma\rho}}F_{\sigma\rho}.
\end{equation}
The dielectric tensor $\chi\indices{_{\mu\nu}^{\alpha\beta}}$ is independently
antisymmetric under index exchange on either $\mu\nu$ or
$\alpha\beta$,
allowing $\mc$ the maximum of 36 independent parameters that can fully relate the six independent components of $\mG$ and $\mF$.
It can be thought of as a 2-form-valued bivector that operates on 2-forms.
In this constitutive relation, the vacuum is uniquely defined as the trivial dielectric such that $\mc_{vac}\mF=\mF$.
The usual permeability, permittivity, and magnetoelectric coupling matrices can be extracted from $\mc$ and interpreted in their usual sense in a locally flat frame, but these traditional matrix quantities intrinsically mix components of $\mc$ with components of the metric even in flat space-time, where, for example, factors of $r$ appear in spherical or cylindrical coordinates.

\subsection{Geometric optics limit}
The wave propagation of light is described by a second order equation, but Maxwell's equations provide two first order equations.
The most straightforward approach to geometric optics is to consider the 1-form potential $\mA=A_{\mu}$, which in a locally flat Minkowski space-time is equivalent to the 4-vector potential $A^{\mu} = (\phi, \vec{A})$.
The field strength tensor $\mF$ is related to $\mA$ by
\begin{equation}
\mF = \ed\mA,
\end{equation}
and the nilpotency of the exterior derivative automatically guarantees the satisfaction of Maxwell's homogeneous equation
\begin{equation}
\ed\mF = \ed^2 \mA = 0,
\end{equation}
while in vacuum a Green's function approach to the solution of the inhomogeneous equation
\begin{equation}
 \ed\star \ed \mA = \mJ
\end{equation}
leads to Jefimenko's equations \cite{Jackson}.

Since we are interested in the geometric optics of light propagation through macroscopically neutral media, we consider
\begin{equation}
 \ed \star \mc \ed \mA = 0.
\end{equation}
Operating on both sides of this equation with the Hodge dual, it may be rewritten in terms of the codifferential of a $k$-form
\begin{equation}
 \delta = (-1)^{k(m+1)-1}\star\ed\star
\end{equation}
as
\begin{equation} \label{Eq:InhomogA}
 \delta \mc\ed\mA = 0.
\end{equation}
While the exterior derivative increases the degree of a differential form by 1, e.g., $\ed$ of a 1-form results in a 2-form, the codifferential decreases the degree by 1; 
so while $\mc\ed\mA$ is a 2-form, $\delta \mc\ed\mA$ is a 1-form. 
The codifferential of a $k$-form can be given an indexed expression in terms of ordinary derivatives and the metric, allowing Eq.~(\ref{Eq:InhomogA}) to be recast as
\begin{equation} \label{Eq:InhomogAindex}
 (\delta \chi\ed\mA)_{\alpha} = g_{\alpha\beta}\frac{1}{\sqrt{|g|}}\partial_{\nu}\left(\sqrt{|g|} g^{\nu\sigma}g^{\beta\rho} \chi\indices{_{\sigma\rho}^{\lambda\kappa}}\partial_{[\lambda}A_{\kappa]}\right) = 0.
\end{equation}

To reach the geometric optics limit we assume a solution of the form
\begin{equation} \label{Eq:PlaneWave}
A_{\mu} = \hat{A}_{\mu}(x^{\rho})e^{i S(x^{\rho})}
\end{equation}
and require that the amplitude $\hat{A}_{\mu}(x^{\rho})$ is slowly varying with respect to the phase $S(x^{\rho})$, subject to the constraint \cite{Post}
\begin{equation}
 \left|\frac{\partial_{\nu}\hat{A}_{\mu}}{\hat{A}_{\mu}}\right| \ll \partial_{\nu}S.
\end{equation}
By furthermore requiring the space-time and the medium to also be slowly varying compared with $S$,
\begin{equation}
  \left|\frac{\partial_{\nu}\chi\indices{_{\mu\alpha}^{\sigma\rho}}}{\chi\indices{_{\mu\alpha}^{\sigma\rho}}}\right| \ll \partial_{\nu}S \quad \mbox{and} \quad 
  \left|\frac{\partial_{\nu}g^{\sigma\rho}}{g^{\sigma\rho}}\right| \ll \partial_{\nu}S,
\end{equation}
one finds that the principal part of Eq.~(\ref{Eq:InhomogAindex}), called the eikonal equation, is
\begin{equation} \label{Eq:Eikonal}
 g^{\nu\sigma}\chi\indices{_{\sigma\alpha}^{\lambda\kappa}}(\partial_{\nu}S)(\partial_{\lambda}S)\hat{A}_{\kappa} = g^{\nu\sigma}\chi\indices{_{\sigma\alpha}^{\lambda\kappa}}p_{\nu}p_{\lambda}\hat{A}_{\kappa} = 0,
\end{equation}
where we write $p_{\nu} = \partial_{\nu}S$ (or the index-free version $\bm{p}=\ed S$).
In other terminology, this is equivalent to a WKB approximation, where one assumes a solution of the form $\mA = \hat{\mA}e^{i\frac{S}{\lambda}}$ and retains the leading order terms of Eq.~(\ref{Eq:InhomogAindex}) in the limit $\lambda\to 0$ \cite{BornWolf}.

Writing Eq.~(\ref{Eq:Eikonal}) as
\begin{equation}
  X\indices{_{\alpha}^{\kappa}}\hat{A}_{\kappa} = 0
\end{equation}
and thinking of 
\begin{equation} \label{Eq:Xmatrix}
  X\indices{_{\alpha}^{\kappa}} = g^{\nu\sigma}\chi\indices{_{\sigma\alpha}^{\lambda\kappa}}p_{\nu}p_{\lambda}
\end{equation}
as a $4\times 4$ matrix, it may be seen that the existence of a nontrivial solution to Eq.~(\ref{Eq:Eikonal}) requires
\begin{equation}
\det(\bm{X}) = 0.
\end{equation}
In fact, this condition is satisfied identically.  
By the antisymmetry of the second set of indices on $\mc$, $\hat{A}_{\mu}\propto p_{\mu}$ is already a trivial solution, so any nontrivial solution resides in the three-dimensional subspace orthogonal to $p_{\mu}$, meaning the matrix is effectively only three dimensional \footnote{The orthogonality of $\mA$ and $\bm{p}$ is also a consequence of the Lorenz condition, the adoption of which is required in conjunction with the assumed plane wave solution Eq.~(\ref{Eq:PlaneWave})}.
There are some different methods for dealing with this (see, for example, Ref.\ \cite{Hehl2002ijmpa}); we offer a purely algebraic argument based on the classical adjugate matrix $\mathrm{adj}(\bm{X})$, defined such that
\begin{equation}
 \bm{X} \mathrm{adj}(\bm{X}) = \det(\bm{X})\bm{I}.
\end{equation} 
If $\bm{X}$ is invertible, then $\mathrm{adj}(\bm{X}) \propto \bm{X}^{-1}$, but $\mathrm{adj}(\bm{X})$ is defined even if  $\bm{X}^{-1}$ does not exist.
Since $\det(\bm{X}) = 0$ identically, then it must be true that $\bm{X}\mathrm{adj}(\bm{X}) = 0$.  Since $\bm{X}$ is nonzero and arbitrary, the subsidiary condition
\begin{equation}
 \mathrm{adj}(\bm{X}) = 0
\end{equation}
must be satisfied.  
Although this is a matrix condition, one may show that it is actually of the form
\begin{equation}
 \mathrm{adj}(\bm{X}) = P (\bm{p}\otimes\bm{p}^{\sharp})
\end{equation}
where $P$ is a polynomial of fourth order in $p_{\mu}$ and $(\bm{p}\otimes\bm{p}^{\sharp})\indices{_{\mu}^{\nu}} = p_{\mu}g^{\nu\alpha}p_{\alpha}$ is a nonzero matrix consisting of simple multiples of components of $p_{\mu}$.  Therefore, the condition for nontrivial solutions to Eq.\ (\ref{Eq:Eikonal}) reduces to the scalar condition
\begin{equation}
P = 0.
\end{equation}
Since we are interested in characterizing the medium with an (possibly psuedo-Finslerian) optical metric, we restrict ourselves to media for which the polynomial $P$ can be factored to
\begin{equation}
 P = H_+ H_- = (a^{\alpha\beta}p_{\alpha}p_{\beta} + \sqrt{F^{\mu\nu\sigma\rho}p_{\mu}p_{\nu}p_{\sigma}p_{\rho}}) (a^{\alpha\beta}p_{\alpha}p_{\beta} - \sqrt{F^{\mu\nu\sigma\rho}p_{\mu}p_{\nu}p_{\sigma}p_{\rho}})=0.
\end{equation}
Such a factorization may require the imposition of some symmetry conditions on $\boldsymbol{\chi}$ that reduce the number of free parameters from 36, but determining the space of allowable $\boldsymbol{\chi}$ is beyond the scope of the present work.  For our purpose it is sufficient to assume that $P$ is factorizable so.

Thus, it can be seen that there are two possible branches, $H_+ =0$ or $H_-=0$, corresponding to two different propagation states of light; i.e.\ the medium displays birefringence for which there exist two independent propagating eigenstates that follow distinct ray trajectories.
The solutions depend not only on the location within the medium, but also on the direction of light propagation.
Since $H_+$ and $H_-$ are quadratic in $p_{\mu}$, each branch furthermore supports two future-directed solutions traveling in opposite directions, e.g.\ left/right or ingoing/outgoing rays.

For each propagation state, the corresponding function $H_{+}(x^{\mu},p_{\mu})$ or $H_{-}(x^{\mu},p_{\mu})$ defines a pseudo-Finslerian structure on the cotangent bundle.
One may then define two associated pseudo-Finslerian \textit{optical metrics},
\begin{equation}
 \gamma_{\pm}^{\alpha\beta} = \frac{\partial^2 H_{\pm}}{\partial p_{\alpha}\partial p_{\beta}}.
\end{equation}
This may be summarized by saying that in the geometric optics limit, there exist two branches of on-shell solutions of Maxwell's equations at each point that must everywhere satisfy the corresponding conditions
\begin{equation}
 H_+ = \frac12\gamma_+^{\alpha\beta}(x^{\mu},p_{\mu})p_{\alpha}p_{\beta} =0, \qquad \mbox{or} \qquad  H_- = \frac12 \gamma_-^{\alpha\beta}(x^{\mu},p_{\mu})p_{\alpha}p_{\beta} =0,
\end{equation}
where we emphasize that $\gamma_{\pm}^{\mu\nu}$ depends on both the position and the wave vector.
The fact that these equations have the same form as the condition for null curves on a vacuum manifold,
\begin{equation}
 \frac12 g^{\alpha\beta}p_{\alpha}p_{\beta}=0
\end{equation}
provides the source of the terminology ``optical metric,'' but it is important to realize that while the optical metric defines the light cone, it is an emergent quantity that serves as an additional structure on the manifold and is not a replacement of the background space-time metric.
Physical observables are still measured with respect to the background space-time metric, and the connection (covariant derivative) is still usually adopted as that which is compatible with the background metric.

From Eq.~(\ref{Eq:Xmatrix}), it is clear that the $\gamma^{\alpha\beta}_{\pm}$ contain contributions from both the background space-time metric and the medium.
The $\gamma^{\alpha\beta}_{\pm}$ fall into one of three categories depending on the properties of the medium:
\begin{enumerate}
\item If the medium parameters are all set to their vacuum values (vanishing media), then $F^{\mu\nu\sigma\rho}=0$ and $a^{\alpha\beta}=\frac12 g^{\alpha\beta}$, and the optical metrics degenerate to the space-time metric $\gamma_+^{\alpha\beta} = \gamma_-^{\alpha\beta} = g^{\alpha\beta}$.  
This shows a) that light in vacuum follows the null curves of the space-time and b) that the space-time metric also serves as the optical metric of the vacuum, which itself can be considered a trivial dielectric.
The fact that the space-time metric serves as the optical metric in vacuum is not surprising since it is the only available structure in the absence of ponderable media.  Ponderable dielectric media provide additional structure that contributes to the optical metric.

\item For impedance matched media, $F^{\mu\nu\sigma\rho}=0$.  In this case, the two optical metrics are everywhere degenerate (no birefringence) but distinct from the background metric, $\gamma_+^{\alpha\beta} = \gamma_-^{\alpha\beta} \neq g^{\alpha\beta}$. 
Since the square root term containing $F^{\mu\nu\sigma\rho}$ vanishes, the degenerate optical metric now has the properties of a pseudo-Riemannian metric instead of a pseudo-Finslerian metric, but it should still be understood as comprised of some combination of the medium parameters and the background space-time in which the medium resides.

\item In all other media, the two optical metrics are typically distinct, $\gamma_+^{\alpha\beta}\neq \gamma_-^{\alpha\beta} \neq g^{\alpha\beta}$, but it is possible that they degenerate for specially chosen incident waves with $p_{\mu}$ such that $F^{\mu\nu\sigma\rho}p_{\mu}p_{\nu}p_{\sigma}p_{\rho} =0$.
Such $p_{\mu}$ correspond to light propagating along an optical axis of the medium.

\end{enumerate}

\section{Kinematics of light in media} \label{Sec:Kinematics}
The fact that the optical metric is not the same as the space-time metric reflects the fact that light in media is generically  nongeodesic and non-null with respect to the background space-time, and the light cone within the medium does not coincide with the light cone of the space-time.
We reiterate that although ray propagation is governed by the optical metric, the structure imposed by the background space-time is still present, and an observer still makes physical measurements of length and angle with respect to $g^{\alpha\beta}$ rather than $\gamma^{\alpha\beta}$.

In this section, we study the kinematical decomposition of a congruence of light in arbitrary dielectric media.
The formalism developed here enables ray tracing through smoothly varying media and allows us to follow the congruence regardless of whether it is null, timelike, or spacelike with respect to the background metric.  

We begin with Hamilton's canonical equations for the congruence, which gives us a general prescription for ray tracing through any media residing in any space-time.
Note that since these equations are derived directly from the geometric optics limit of Maxwell's equations rather than from a Lagrangian, there is no assumption of energy conservation and this formalism may be used for ray tracing through time-dependent media.

Once the congruence has been established, we define a subspace that is transverse to the optical light cone (i.e.\ the light cone of the medium rather than the background light cone).
Physically, for example, this corresponds to looking at the cross sectional slice of a laser beam -- a surface of constant phase.  
We then look at the evolution of this surface as it propagates along the beam to see how it expands or contracts, rotates, or shears.

From here on, we consider only a single propagation eigenstate by choosing either of $H_+$ or $H_-$ and dropping the index so that we have a generic $H$. We have seen that in the geometrical optics limit, on-shell solutions of Maxwell's equations must everywhere satisfy the condition $H=0$.
This means that if light follows some curve $\gamma$ parametrized by $\tau$, then
\begin{equation} \label{Eq:Hderivative}
\frac{dH}{d\tau} = 0
\end{equation}
everywhere along the curve.
Since $H = H(x^{\mu},p_{\mu})$ is a function on the cotangent bundle,
\begin{equation} \label{Eq:Hderivative2}
 \frac{dH}{d\tau} = \left(\frac{d x^{\mu}}{d\tau}\right)\frac{\partial H}{\partial x^{\mu}} + \left(\frac{d p_{\mu}}{d\tau}\right)\frac{\partial H}{\partial p_{\mu}} = \dot{x}^{\mu} \frac{\partial H}{\partial x^{\mu}} + \dot{p}_{\mu} \frac{\partial H}{\partial p_{\mu}} = 0,
\end{equation}
which implies Hamilton's canonical equations
\begin{equation} \label{Eq:HamiltonsEqs}
 \dot{x}^{\mu} = \frac{\partial H}{\partial p_{\mu}},\quad \dot{p}_{\mu} = -\frac{\partial H}{\partial x^{\mu}}.
\end{equation}
Let
\begin{equation}
 s^{\mu} = \dot{x}^{\mu}
\end{equation}
denote the tangent to $\gamma$.
Since $\gamma$ is not necessarily geodesic, we have nontrivial acceleration $a^{\alpha}$,
\begin{equation} \label{Eq:acceleration}
 s^{\beta}s\indices{^{\alpha}_{;\beta}} = a^{\alpha}.
\end{equation}

In a vacuum space-time it is true that $g_{\mu\nu}s^{\mu}=s_{\nu}=p_{\nu}$, but  
this is no longer true in media.  
Instead, in the presence of the optical metric $\gamma^{\alpha\beta}$ Hamilton's equations result in $s^{\mu}=\gamma^{\mu\nu}p_{\nu}$, and subsequently it is no longer true that $g_{\alpha\beta}s^{\alpha}s^{\beta}=0$.
An observer making measurements relative to the space-time metric will find $s^{\mu}\neq g^{\mu\nu}p_{\mu}$, and will be lead to the conclusion that the Poynting vector does not always coincide with the direction of wave propagation inside anisotropic media, but from Hamilton's equations and the form of $H$ it is always true that $s^{\alpha}p_{\alpha}=0$.

It is worth noting that there is an arbitrariness in the definition of $H$ since if $H(x^{\mu},p_{\mu})=0$ then $C(x^{\mu})H(x^{\mu},p_{\mu})=0$ with monotonic $C(x^{\mu})\neq 0$ will also be satisfied.  
This is equivalent to a rescaling of the parameter along $\gamma$. The curve $\gamma$ therefore represents an equivalence class of curves through the same space-time points but visited at different parameter values.

The preceding statements would benefit from some clarification.  A solution to Maxwell's equations in the geometric optics limit requires information about both the position $x^{\mu}$ and the wave (co)vector $p_{\mu}$.
The parametrized curve $\gamma$ is a curve $\gamma: \mathbb{R} \to T^*M$ in the cotangent bundle given by
\begin{equation}
 \gamma(\tau) = (x^{\mu}(\tau),p_{\mu}(\tau)).
\end{equation}
Thus, the space of solutions to Maxwell's equations is the eight-dimensional cotangent bundle (i.e., the phase space), and Hamilton's Eqs.~(\ref{Eq:HamiltonsEqs}) provide eight equations for the eight components of the solution curve.
On the other hand, the physical ray trajectory, $\tilde{\gamma}$, is a curve on the space-time manifold $M$ found by projecting from the cotangent bundle with
\begin{equation}
 \begin{aligned}
 & \pi: T^*M \to M\\
 \tilde{\gamma}(\tau) = \ & \pi(\gamma(\tau)) = x^{\mu}(\tau).
 \end{aligned}
\end{equation}
Since $\pi$ simply selects the $x^{\mu}$ components of a point $q\in T^*M$, the projection of $\gamma(\tau)$ into $\tilde{\gamma}(\tau)$ is a relatively trivial matter, but confusion can arise if the distinction is not pointed out.  
For example, the tangent to a curve in $T^*M$ has components over $\{\partial/\partial p_{\mu}\}$ and so we see derivatives with respect to $p_{\mu}$ in Eq. (\ref{Eq:Hderivative2}), whereas the derivative of a function $F(x^{\mu})$ on $M$ along $\tilde{\gamma}$ would only be of the form $s^{\mu}F_{,\mu}$.
Similarly, denoting $s^{\mu}$ as the ``tangent to $\gamma$'' as mentioned just before Eq.~(\ref{Eq:acceleration}) is misleading, as it is actually the projection of the tangent to $\gamma$, or equivalently the tangent of the projected curve $\tilde{\gamma}$.
The same is true for the acceleration of Eq.~(\ref{Eq:acceleration}), which is actually that of $\tilde{\gamma}$.

\subsection{Jacobi fields}
With a congruence of rays representing solutions to Eqs.~(\ref{Eq:HamiltonsEqs}) in hand, we look for a characterization of the subspace transverse to the congruence.  
To reiterate, we are now looking at a congruence of ray trajectories $\bar{\gamma}(\tau)\subset M$ that are the projection onto $M$ of a congruence of solution curves $\gamma(\tau)\subset T^*M$, and we choose one particular curve $\tilde{\gamma}$ to act as a reference curve.
We take the chosen reference curve to be the particular solution 
\begin{equation}
  (s^{\mu},p_{\mu}) = (u^{\mu},k_{\mu}),
\end{equation}
to Eqs.~(\ref{Eq:HamiltonsEqs}), and thus $\bm{u}$ is tangent to the chosen curve $\tilde{\gamma}(\tau)$.

The first step is to find a set of vector fields that can be tracked along the congruence and which will provide a measure for changes along the congruence.
It is always possible to find a set of vector fields $\{\boldsymbol{\xi}\}$ that are Lie invariant along the curve such that
\begin{equation} \label{Eq:NullCurveJacobiField}
 \mathcal{L}_{\bm{u}}\boldsymbol{\xi} = u^{\beta}\xi\indices{^{\alpha}_{;\beta}} - \xi^{\beta}u\indices{^{\alpha}_{;\beta}} = 0.
\end{equation}
Such vector fields are called Jacobi fields.
It may be shown that at a point $p$ on the reference curve $\tilde{\gamma}$, the vector $\boldsymbol{\xi}(p)$ is tangent to a curve (not a member of the congruence) that connects $p$ to a point $q$ on a nearby curve in the congruence \cite{Poisson}.
Physically, the Jacobi field appears to fill the role of characterizing nearby geodesics, but note that the tangent vector field $\bm{u}$ is itself Lie invariant along $\tilde{\gamma}$, so not every Jacobi field fulfills the desired role of pointing to a nearby curve in the congruence, and what we want is a Jacobi field that points transversely to a nearby curve (at least initially). 

For timelike congruences in vacuum, one typically enforces the choice $\xi^{\beta}u_{\beta}=0$ to ensure that $\boldsymbol{\xi}$ is transverse, but since $u^{\alpha}k_{\alpha}=0$ this condition is, by itself, insufficient to guarantee that $\boldsymbol{\xi}$ does not have a component along $\bm{u}$.
Consequently, at any point $p$ on $\gamma$, one may wish to decompose $\boldsymbol{\xi}$ into a piece that is parallel to $\bm{u}$ and a piece that is orthogonal to $\bm{u}$,
\begin{equation}
\xi^{\alpha} = Au^{\alpha} + \bar{\xi}^{\alpha},
\end{equation}
recognizing that only the orthogonal part of $\boldsymbol{\xi}$ provides the desired information about the transverse behavior of the congruence.
But as shown below, this is only a partial decomposition of $\boldsymbol{\xi}$ that must be augmented by a missing component.

\subsection{Projection operator}
In vacuum, the self-orthogonality of a null curve, $u^{\mu}u_{\mu}$=0, makes it difficult to identify the truly transverse component of $\boldsymbol{\xi}$.  To do so requires an operator that projects a vector into the subspace transverse to $\bm{u}$.
As mentioned at the beginning of this section, the cross sectional measurement of a laser beam is made at some instant of time at a fixed position along the beam length; e.g., for a beam propagating in the $z$-direction, one would make a measurement in the $xy$ plane at some fixed $z$.
The transverse subspace is therefore two dimensional, while the complimentary space (the light cone) is also two dimensional.  The vector $\bm{u}$ does not, by itself, span the light cone, and we require a second null vector to help define the transverse subspace via a projection operator.

In the literature, dealing with null congruences in vacuum, this ``auxiliary null vector'' is usually introduced arbitrarily, with the arbitrariness justified \textit{post facto} because it drops out of the final answer. 
However, this turns out to no longer be true in the more general setting pursued here, and we will be forced to seek a canonical construction of the auxiliary vector that is based on some physically meaningful quantity.
The details of this construction will be deferred until Sec.\ \ref{Sec:Raychaudhuri}.
For now, we assume the existence of two future-directed lightlike pairs $(u^{\alpha},k_{\alpha})$, and $(v^{\alpha},\ell_{\alpha})$ at every point, that satisfy 
\begin{equation} \label{Eq:NullConditions}
u^{\alpha}k_{\alpha}=0, \quad v^{\alpha}\ell_{\alpha}=0, \quad u^{\alpha}\ell_{\alpha}\neq 0, \quad v^{\alpha}k_{\alpha} \neq 0.
\end{equation}
Defining the projection operator
 \begin{equation} \label{Eq:NullProjection}
 h^{\alpha}_{\beta} = \delta^{\alpha}_{\beta} - \frac{\ell_{\alpha}u^{\beta}}{\ell_{\sigma}u^{\sigma}} - \frac{k_{\alpha}v^{\beta}}{k_{\sigma}v^{\sigma}},
 \end{equation}
one may readily demonstrate that $u^{\beta}h^{\alpha}_{\beta}= v^{\beta}h^{\alpha}_{\beta} = k_{\alpha}h^{\alpha}_{\beta} = \ell_{\alpha}h^{\alpha}_{\beta} = 0$ and that
 \begin{equation} \label{Eq:ProjectorMultiplication}
 h^{\alpha}_{\beta} h^{\beta}_{\mu} = h^{\alpha}_{\mu}.
 \end{equation}
Thus, if $\bm{u}$ and $\bm{v}$ span the light cone at a point $p\in M$, then for any any arbitrary vector $\bm{c}$, $\bm{h}(\bm{c})$ is orthogonal to both $\bm{u}$ and $\bm{v}$ and therefore lies in the subspace transverse to the light cone.

The trace
\begin{equation}
h^{\alpha}_{\alpha} =2
\end{equation}
indicates that the subspace transverse to both $\bm{u}$ and $\bm{v}$ is still two dimensional, despite the fact that the trajectories followed by light may in fact be timelike or spacelike with respect to the background metric.
One can already see some of the issues that will be faced in dealing with this definition of the projection operator:
For null geodesics in vacuum the usual technique is to set the denominators appearing in Eq.~(\ref{Eq:NullProjection}) to -1, but since in general media the optical metric depends on $p_{\mu}$, the denominators $(\ell_{\sigma}u^{\sigma})=(\gamma_{\sigma\rho}(\boldsymbol{\ell})v^{\rho}u^{\sigma})$ and $(k_{\sigma}v^{\sigma})=(\gamma_{\sigma\rho}(\bm{k})u^{\rho}v^{\sigma})$ are not even the same, and it is unclear whether, and under what conditions, they may be made constant, and whether they can be simultaneously constant.

Since scalar factors like $(\ell_{\sigma}u^{\sigma})$ or $(k_{\sigma}v^{\sigma})$ appear frequently, we suppress the indices and write $(\ell\cdot u)$ or $(k\cdot v)$.
In all such factors it should be understood that a vectorial quantity is being contracted with a covectorial quantity and does not involve the metric.

\subsection{Transverse evolution of a congruence in media}
With this more detailed understanding of the null structure at a point, we can see that the most general decomposition of any vector is of the form
\begin{equation} \label{Eq:XiDecomposition}
\xi^{\alpha} = Au^{\alpha} + Bv^{\alpha} + \bar{\xi}^{\alpha}.
\end{equation}
Since $h^{\alpha}_{\beta}u^{\beta}=h^{\alpha}_{\beta}v^{\beta}=0$ and $\bar{\xi}^{\alpha}$ is defined to lie in the transverse subspace, it follows that
\begin{equation}
\bar{\xi}^{\alpha} = h^{\alpha}_{\beta}\xi^{\beta}.
\end{equation}
The sought-after transverse kinematics of the congruence is a description of the evolution of $\bar{\boldsymbol{\xi}}$ along the chosen reference curve $\tilde{\gamma}$.  If $\tau$ parametrizes the curve, then near the point $\tau_0$
\begin{equation} \label{Eq:TaylorExp}
 \boldsymbol{\bar{\xi}}(\tau) = \boldsymbol{\bar{\xi}}(\tau_0) + \left.\frac{d\boldsymbol{\bar{\xi}}}{d\tau}\right|_{\tau_0}\Delta \tau + O(\Delta \tau)^2 .
\end{equation}
In components we have
\begin{equation}
\left(\frac{d\bar{\boldsymbol{\xi}}}{d\tau}\right)^{\alpha} = \dot{\bar{\xi}}\indices{^{\alpha}} = u^{\beta}\bar{\xi}\indices{^{\alpha}_{;\beta}} = u^{\beta}\left(h^{\alpha}_{\mu}\xi^{\mu}\right)_{;\beta} = u^{\beta}h^{\alpha}_{\mu;\beta}\xi^{\mu} + h^{\alpha}_{\mu}u^{\beta}\xi\indices{^{\mu}_{;\beta}}.
\end{equation}
Since $\boldsymbol{\xi}$ is a Jacobi field, Eq.\ (\ref{Eq:NullCurveJacobiField}) implies 
\begin{equation}
u^{\beta}\xi\indices{^{\mu}_{;\beta}} = \xi^{\beta}u\indices{^{\mu}_{;\beta}},
\end{equation}
whence
\begin{equation}
\dot{\bar{\xi}}\indices{^{\alpha}} = \left(u^{\sigma}h^{\alpha}_{\beta;\sigma} + h^{\alpha}_{\mu}u\indices{^{\mu}_{;\beta}}\right)\xi^{\beta}.
\end{equation}
Note that since $h^{\alpha}_{\mu}u^{\mu}=0$, then $(h^{\alpha}_{\mu}u^{\mu})_{;\beta}=0$ implies
\begin{equation}
 h^{\alpha}_{\mu}u\indices{^{\mu}_{;\beta}} = -u^{\sigma}h^{\alpha}_{\sigma;\beta},
\end{equation}
so we have
\begin{equation} \label{Eq:TransEvolution1}
 \dot{\bar{\xi}}\indices{^{\alpha}} = u^{\sigma}\left(h^{\alpha}_{\beta;\sigma} -  h^{\alpha}_{\sigma;\beta}\right)\xi^{\beta} = u^{\sigma}\left(h^{\alpha}_{\beta;\sigma} -  h^{\alpha}_{\sigma;\beta}\right)\left(Au^{\beta}+Bv^{\beta}+ \bar{\xi}^{\beta}\right).
\end{equation}
The antisymmetrization immediately kills any terms dependent on $A$, but the other terms must be examined more carefully.
It follows from Eq.~(\ref{Eq:NullProjection}) that
\begin{multline}
h^{\alpha}_{\beta;\sigma} = (\ell\cdot u)^{-2}(\ell_{\tau;\sigma}u^{\tau}+\ell_{\tau}u\indices{^{\tau}_{;\sigma}})\ell_{\beta}u^{\alpha} - (\ell\cdot u)^{-1}(\ell_{\beta;\sigma}u^{\alpha}+\ell_{\beta}u\indices{^{\alpha}_{;\sigma}})+ \\
(k\cdot v)^{-2}(k_{\tau;\sigma}v^{\tau}+k_{\tau}v\indices{^{\tau}_{;\sigma}})k_{\beta}v^{\alpha} - (k\cdot v)^{-1}(k_{\beta;\sigma}v^{\alpha}+k_{\beta}v\indices{^{\alpha}_{;\sigma}}).
\end{multline}
Calculating the $B$ term in Eq.~(\ref{Eq:TransEvolution1}), one finds
\begin{equation} \label{Eq:Bterm1}
    B u^{\sigma}v^{\beta}(h^{\alpha}_{\beta;\sigma}-h^{\alpha}_{\sigma;\beta})  = B\left[ v^{\beta}u\indices{^{\alpha}_{;\beta}} - u^{\beta}v\indices{^{\alpha}_{;\beta}} - (\ell\cdot u)^{-1}u^{\alpha}v^{\beta}(\ell_{\tau}u\indices{^{\tau}_{;\beta}} + \ell_{\beta;\tau}u^{\tau})  + 
    (k\cdot v)^{-1}v^{\alpha}u^{\beta}(k_{\tau}v\indices{^{\tau}_{;\beta}} + k_{\beta;\tau}v^{\tau}) \right].
\end{equation}
By Eq.~(\ref{Eq:NullConditions}), $u^{\beta}k_{\beta;\tau} = -k_{\tau}u\indices{^{\tau}_{;\beta}}$ and $v^{\beta}\ell_{\beta;\tau} = -\ell_{\tau}v\indices{^{\tau}_{;\beta}}$, allowing Eq.~(\ref{Eq:Bterm1}) to be recast as
\begin{equation}
 \begin{aligned}
B u^{\sigma}v^{\beta}(h^{\alpha}_{\beta;\sigma}-h^{\alpha}_{\sigma;\beta})  & = B\left[v^{\beta}(u\indices{^{\alpha}_{;\beta}} - (\ell\cdot u)^{-1} u^{\alpha}\ell_{\tau} u\indices{^{\tau}_{;\beta}} -(k\cdot v)^{-1} v^{\alpha}k_{\tau} u\indices{^{\tau}_{;\beta}})  \right. \\
 & \qquad \left. - u^{\beta}(v\indices{^{\alpha}_{;\beta}} - (\ell\cdot u)^{-1} u^{\alpha}\ell_{\tau} v\indices{^{\tau}_{;\beta}} -(k\cdot v)^{-1} v^{\alpha}k_{\tau} v\indices{^{\tau}_{;\beta}})  \right] \\
 & = B(v^{\beta}h^{\tau}_{\beta}u\indices{^{\alpha}_{;\tau}} - u^{\beta}h^{\tau}_{\beta}v\indices{^{\alpha}_{;\tau}}) \\
 & = 0.
 \end{aligned}
\end{equation}

So far we have found that, fortunately, the evolution of $\bar{\boldsymbol{\xi}}$ does not depend on either of the undetermined parameters $A$ or $B$ in Eq.~(\ref{Eq:XiDecomposition}).
Turning attention to the $\bar{\xi}^{\beta}$ term of Eq.~(\ref{Eq:TransEvolution1}), and using the convolution property of $h^{\alpha}_{\beta}$ given in Eq.\ (\ref{Eq:ProjectorMultiplication}) to write $\bar{\xi}^{\alpha} = h^{\alpha}_{\omega}\xi^{\omega} = h^{\alpha}_{\lambda}h^{\lambda}_{\omega}\xi^{\omega} = h^{\alpha}_{\lambda}\bar{\xi}^{\lambda}$, one finds
\begin{equation}
 \begin{aligned} 
 \dot{\bar{\xi}}\indices{^{\alpha}} = u^{\sigma}\left(h^{\alpha}_{\beta;\sigma} -  h^{\alpha}_{\sigma;\beta}\right)h^{\beta}_{\lambda}\bar{\xi}^{\lambda} & 
 = (\ell\cdot u)^{-2}\left[u^{\sigma}\left(\ell_{\rho;\sigma}u^{\rho} + \ell_{\rho}u\indices{^{\rho}_{;\sigma}}\right)(\ell_{\beta}h^{\beta}_{\lambda})u^{\alpha} - \left(\ell_{\rho;\beta}u^{\rho} + \ell_{\rho}u\indices{^{\rho}_{;\beta}}\right)h^{\beta}_{\lambda}(u^{\sigma}\ell_{\sigma})u^{\alpha} \right] \bar{\xi}^{\lambda}
 \\ & - (\ell\cdot u)^{-1} \left[u^{\alpha}u^{\sigma}\left(\ell_{\beta;\sigma}-\ell_{\sigma;\beta}\right)h^{\beta}_{\lambda} + (\ell_{\beta}h^{\beta}_{\lambda})u^{\sigma}u\indices{^{\alpha}_{;\sigma}} - (u^{\sigma}\ell_{\sigma})u\indices{^{\alpha}_{;\beta}}h^{\beta}_{\lambda} \right] \bar{\xi}^{\lambda}
\\ & + (k\cdot v)^{-2}\left[u^{\sigma}\left(k_{\rho;\sigma}v^{\rho} + k_{\rho}v\indices{^{\rho}_{;\sigma}}\right)(k_{\beta}h^{\beta}_{\lambda})v^{\alpha} - \left(k_{\rho;\beta}v^{\rho} + k_{\rho}v\indices{^{\rho}_{;\beta}}\right)h^{\beta}_{\lambda}(u^{\sigma}k_{\sigma})v^{\alpha} \right] \bar{\xi}^{\lambda}
\\ & - (k\cdot v)^{-1} \left[v^{\alpha}u^{\sigma}\left(k_{\beta;\sigma}-k_{\sigma;\beta}\right)h^{\beta}_{\lambda} + (k_{\beta}h^{\beta}_{\lambda})u^{\sigma}v\indices{^{\alpha}_{;\sigma}} - (u^{\sigma}k_{\sigma})v\indices{^{\alpha}_{;\beta}}h^{\beta}_{\lambda} \right] \bar{\xi}^{\lambda}.
 \end{aligned}
\end{equation}
By the action of the projection operator on $\ell_{\alpha}$ and $k_{\alpha}$, together with the fact that $u^{\sigma}k_{\sigma;\beta} = -k_{\sigma}u\indices{^{\sigma}_{;\beta}}$, 
\begin{equation}
 \begin{aligned}
  \dot{\bar{\xi}}\indices{^{\alpha}} & = \left[u\indices{^{\alpha}_{;\beta}} - (\ell\cdot u)^{-1}\ell_{\rho}u^{\alpha}u\indices{^{\rho}_{;\beta}} - (k\cdot v)^{-1}k_{\rho}v^{\alpha}u\indices{^{\rho}_{;\beta}}   -(\ell\cdot u)^{-1}u^{\alpha}u^{\sigma}\ell_{\beta;\sigma} -(k\cdot v)^{-1}v^{\alpha}u^{\sigma}k_{\beta;\sigma}\right]h^{\beta}_{\lambda}\bar{\xi}^{\lambda}
  \\ & = h^{\alpha}_{\rho}u\indices{^{\rho}_{;\beta}}h^{\beta}_{\lambda}\bar{\xi}^{\lambda} -\left[ (\ell\cdot u)^{-1}u^{\alpha}\dot{\ell}_{\beta} +(k\cdot v)^{-1}v^{\alpha}\dot{k}_{\beta}\right]h^{\beta}_{\lambda}\bar{\xi}^{\lambda}.
 \end{aligned}
\end{equation}
It may here be seen that $\dot{\bar{\boldsymbol{\xi}}}$ has terms proportional to $\bm{u}$ and $\bm{v}$, so even though $\bar{\boldsymbol{\xi}}$ is initially transverse it may develop components along $\bm{u}$ and $\bm{v}$.
Note that the second term may be written
\begin{equation}
-\left[ (\ell\cdot u)^{-1}u^{\alpha}\dot{\ell}_{\beta} +(k\cdot v)^{-1}v^{\alpha}\dot{k}_{\beta}\right]h^{\beta}_{\lambda}\bar{\xi}^{\lambda} = (\delta^{\alpha}_{\beta}-h^{\alpha}_{\beta})u^{\sigma}h^{\beta}_{\lambda;\sigma}\bar{\xi}^{\lambda}
\end{equation}
showing conclusively that there are no additional transverse components hidden in this term since it is annihilated by $h^{\mu}_{\alpha}$.
Finally, we express the evolution of $\bar{\xi}^{\alpha}$ by
\begin{equation} \label{Eq:xiEvolution}
 \dot{\bar{\xi}}^{\alpha} = B\indices{^{\alpha}_{\lambda}}\bar{\xi}^{\lambda},
\end{equation}
with
\begin{equation} \label{Eq:B}
 B\indices{^{\alpha}_{\lambda}} = h^{\alpha}_{\rho}u\indices{^{\rho}_{;\beta}}h^{\beta}_{\lambda} +  (\delta^{\alpha}_{\beta}-h^{\alpha}_{\beta})u^{\sigma}h^{\beta}_{\lambda;\sigma}.
\end{equation}

In general, $\bm{B}$ may be decomposed into trace and traceless parts.  
Taking the trace, one finds it comes entirely from the purely transverse part of $\bm{B}$,
\begin{equation} \label{Eq:Theta}
 \Theta = B\indices{^{\alpha}_{\alpha}} = h^{\alpha}_{\rho}u\indices{^{\rho}_{;\beta}}h^{\beta}_{\alpha}.
\end{equation}
One might be tempted to represent the trace component of $\bm{B}$ as something proportional to $\Theta\, \delta^{\alpha}_{\lambda}$, but in general $\delta^{\alpha}_{\lambda}$ does not lie in the transverse subspace, so to ensure that the trace component is correctly represented as lying entirely in the transverse subspace we project $h^{\alpha}_{\sigma}\delta^{\sigma}_{\rho}h^{\rho}_{\lambda} = h^{\alpha}_{\lambda}$ to write the decomposition
\begin{equation}
 B\indices{^{\alpha}_{\lambda}} = \frac{1}{2} \Theta\, h^{\alpha}_{\lambda} + \hat{B}\indices{^{\alpha}_{\lambda}} + (\delta^{\alpha}_{\beta}-h^{\alpha}_{\beta})u^{\sigma}h^{\beta}_{\lambda;\sigma},
\end{equation}
where the factor of $\tfrac12$ is explained by recalling that $h^{\alpha}_{\alpha}=2$.
 
In the literature, the transverse traceless part, $\hat{\bm{B}}$, is usually further reduced to its symmetric and antisymmetric parts, of which the actions on $\bar{\bm{\xi}}$ through Eq.~(\ref{Eq:xiEvolution}) are understood as rotation and shear.
In our more generalized setting, discussions of symmetry beg the question ``symmetric with respect to what?''
For a tensor like $g_{\mu\nu}$, there is a well-defined meaning of symmetry, which is that given two vectors $a^{\mu}$ and $b^{\mu}$, $g_{\mu\nu}a^{\mu}b^{\nu} = g_{\mu\nu}b^{\mu}a^{\nu}$, or written another way, that $\bm{g}(\bm{a},\bm{b})=\bm{g}(\bm{b},\bm{a})$.
But a mixed tensor like $B\indices{^{\alpha}_{\lambda}}$ accepts two different types of objects as its arguments which cannot simply be interchanged.  
Given $\bm{B}(\bm{k},\bm{u}) = B\indices{^{\alpha}_{\lambda}}k_{\alpha}u^{\lambda}$, there is no meaning to the interchange $\bm{B}(\bm{k},\bm{u})\to \bm{B}(\bm{u},\bm{k})$, because the first argument of $\bm{B}(\cdot\, ,\cdot)$ must be a covector rather than a vector, while the second must be a vector rather than a covector.
Any measure of the symmetry of $\bm{B}$ must therefore be made relative to some other known tensor that provides a map between a vector space and its dual.  
The obvious choice is the metric tensor, whence the symmetry or antisymmetry of $B_{\alpha\beta}=g_{\alpha\sigma}B\indices{^{\sigma}_{\beta}}$ has a well-defined meaning.
Although we may seem to belabor this question here, the point is that we now have at our disposal a second natural choice of measure -- the optical metric.

Framed another way, since $\bm{B}$ is the infinitesimal transformation of a real 4-vector $\bar{\boldsymbol{\xi}}_p \in T_p(M) \sim \mathcal{M}$ (where $\mathcal{M}$ denotes flat Minkowski space-time), the most natural interpretation seems to be that $\bm{B}$ is an element of $\mathfrak{o}(1,3)$, the Lie algebra of O(1,3) -- the group of orthogonal transformations. 
After separating out the trace, the remaining traceless parts must lie in the subgroup $\mathfrak{so}(1,3)$, the Lie algebra of SO(1,3) -- the group of norm-preserving orthogonal transformations where the norm is given by the background metric $\bm{g}$.
But again, one could instead prefer to define a group of norm-preserving transformations with respect to the optical metric rather than the background metric, and since the optical metric reduces to the background metric in vacuum there is no inconsistency in this choice.

We argue that an observer continues to make measurements with respect to the background metric and that this is consistent with real-world practice.
Thus, we suggest that the traceless parts of $\bm{B}$ are most naturally decomposed with respect to the generators of $\mathfrak{so}(1,3)$.
The transverse traceless part $\hat{\bm{B}}$ will be decomposed with respect to those generators that lie in the purely spatial transverse space and will therefore correspond to rotation and shear transformations in this plane.
The complimentary, or longitudinal, space is not purely spatial, and therefore the decomposition of the longitudinal traceless part, $(\delta^{\alpha}_{\beta}-h^{\alpha}_{\beta})u^{\sigma}h^{\beta}_{\lambda;\sigma}$ will include generators corresponding to boosts.  These boosts correspond to the fact that $\bar{\boldsymbol{\xi}}$ can acquire a component along $\bm{u}$, meaning that different parts of a phase front can travel at different speeds within the medium, such that a single phase front intercepted at a distant detector is absorbed over some finite window of time.

\section{generalized Raychaudhuri equation in media} \label{Sec:Raychaudhuri}
%

The Raychaudhuri equation tells us about the evolution of the expansion along the congruence, $\dot{\Theta}$, which essentially tells us how the focus of a beam changes along its length, and is instrumental in the proof of focusing and singularity theorems \cite{HawkingEllis}.
In curved vacuum space-times, the Raychaudhuri equation for null geodesic congruences is
\begin{equation} \label{Eq:VacRaychaudhuri}
 \dot{\Theta} = a\indices{^{\rho}_{;\rho}} - \frac12 \Theta^2 - \hat{B}{^{\rho}_{\tau}}\hat{B}{^{\tau}_{\rho}} - R\indices{^{\rho}_{\lambda\rho\tau}}u^{\lambda}u^{\tau}.
\end{equation}
This version of the Raychaudhuri equation admits the possibility of nonaffine parametrization of the curve, in which case $\bm{a} = \kappa \bm{u}$ and the term $a\indices{^{\rho}_{;\rho}}$ contributes $\kappa\Theta$.
The presence of the Riemann tensor in the Raychaudhuri equation means that it is possible to use congruences as a probe for curvature.

In the spirit of adapting tools from general relativity for use in optics, this section will generalize the Raychaudhuri equation to light propagating through dielectric media residing in a curved background space-time.
In particular, the idea of an equivalence between curved space-times and dielectric media suggests that the generalized Raychaudhuri equation should contain a tensor term constructed out of the optical metric and its derivatives, similar to the Riemann tensor term in Eq.~(\ref{Eq:VacRaychaudhuri}).
The existence of such a tensor would be quite useful for the analysis of light propagation in media.
However, while we do indeed find that the generalized equation is modified through the addition of terms proportional to the optical metric and its derivatives, these do not occur in such a way that they might be identified as an effective Riemann or curvature tensor.

Begin by taking the derivative of $\Theta$ along the curve, which from Eq.~(\ref{Eq:Theta}) is
\begin{equation}
 \begin{aligned}
 \dot{\Theta} & = \frac{d}{d\tau} \left(h^{\mu}_{\rho}u\indices{^{\rho}_{;\beta}}h^{\beta}_{\mu}\right) \\
 & = h^{\mu}_{\rho}\frac{d}{d\tau}\left(u\indices{^{\rho}_{;\beta}}\right)h^{\beta}_{\mu}+ \frac{d}{d\tau}\left(h^{\mu}_{\rho}h^{\beta}_{\mu}\right) u\indices{^{\rho}_{;\beta}} = h^{\mu}_{\rho}(u^{\tau}u\indices{^{\rho}_{;\beta\tau}})h^{\beta}_{\mu} +  \dot{h}^{\beta}_{\rho}u\indices{^{\rho}_{;\beta}}\\
 & = h^{\mu}_{\rho}\left(u^{\tau}u\indices{^{\rho}_{;\tau\beta}} - R\indices{^{\rho}_{\lambda\beta\tau}}u^{\lambda}u^{\tau} \right)h^{\beta}_{\mu} +  \dot{h}^{\beta}_{\rho} u\indices{^{\rho}_{;\beta}} \\
 & = h^{\mu}_{\rho}\left(a\indices{^{\rho}_{;\beta}} - u\indices{^{\rho}_{;\tau}}u\indices{^{\tau}_{;\beta}} - R\indices{^{\rho}_{\lambda\beta\tau}}u^{\lambda}u^{\tau} \right)h^{\beta}_{\mu} +   \dot{h}^{\beta}_{\rho}u\indices{^{\rho}_{;\beta}}\\
 & =  h^{\beta}_{\rho}a\indices{^{\rho}_{;\beta}} - h^{\mu}_{\rho}u\indices{^{\rho}_{;\tau}}u\indices{^{\tau}_{;\beta}}h^{\beta}_{\mu} - h^{\beta}_{\rho}R\indices{^{\rho}_{\lambda\beta\tau}}u^{\lambda}u^{\tau} + \dot{h}^{\beta}_{\rho}u\indices{^{\rho}_{;\beta}}.
 \end{aligned}
\end{equation}
Our strategy is to write this as the usual Raychaudhuri equation plus some additional terms that arise as a result of the presence of the medium.
With a little bit of adding and subtracting, we can rewrite
\begin{equation} 
 \begin{aligned}
 h^{\mu}_{\rho}u\indices{^{\rho}_{;\tau}}u\indices{^{\tau}_{;\beta}}h^{\beta}_{\mu} 
 & = (h^{\mu}_{\rho}u\indices{^{\rho}_{;\tau}}h^{\tau}_{\nu}) (h^{\nu}_{\sigma}u\indices{^{\sigma}_{;\beta}}h^{\beta}_{\mu}) 
 + h^{\mu}_{\rho}u\indices{^{\rho}_{;\tau}}(\delta^{\tau}_{\nu}-h^{\tau}_{\nu}) (\delta^{\nu}_{\sigma}-h^{\nu}_{\sigma})u\indices{^{\sigma}_{;\beta}}h^{\beta}_{\mu} \\
 & = B\indices{^{\mu}_{\nu}}B\indices{^{\nu}_{\mu}} + h^{\mu}_{\rho}u\indices{^{\rho}_{;\tau}}(\delta^{\tau}_{\nu}-h^{\tau}_{\nu}) (\delta^{\nu}_{\sigma}-h^{\nu}_{\sigma})u\indices{^{\sigma}_{;\beta}}h^{\beta}_{\mu}
 \end{aligned}
\end{equation}
and expand the Riemann tensor term to
\begin{equation}
  h^{\beta}_{\rho}R\indices{^{\rho}_{\lambda\beta\tau}}u^{\lambda}u^{\tau} = R\indices{^{\rho}_{\lambda\rho\tau}}u^{\lambda}u^{\tau} - (k\cdot v)^{-1} k_{\rho}R\indices{^{\rho}_{\lambda\beta\tau}}u^{\lambda}v^{\beta}u^{\tau}.
\end{equation}
We therefore may write
\begin{multline} \label{Eq:ThetaDot1}
\dot{\Theta} =  
- B\indices{^{\mu}_{\nu}}B\indices{^{\nu}_{\mu}} 
- R\indices{^{\rho}_{\lambda\rho\tau}}u^{\lambda}u^{\tau} 
+ h^{\beta}_{\rho}a\indices{^{\rho}_{;\beta}}
+ (k\cdot v)^{-1} k_{\rho}R\indices{^{\rho}_{\lambda\beta\tau}}u^{\lambda}v^{\beta}u^{\tau} 
-h^{\mu}_{\rho}u\indices{^{\rho}_{;\tau}} (\delta^{\tau}_{\nu}-h^{\tau}_{\nu})
(\delta^{\nu}_{\sigma}-h^{\nu}_{\sigma}) u\indices{^{\sigma}_{;\beta}}h^{\beta}_{\mu}
+ \dot{h}^{\beta}_{\rho}u\indices{^{\rho}_{;\beta}}
\end{multline}
Consider the final two terms. 
Substituting Eq.\ (\ref{Eq:NullProjection}) for the projection operator, one finds
\begin{multline}
h^{\mu}_{\rho}u\indices{^{\rho}_{;\tau}} (\delta^{\tau}_{\nu}-h^{\tau}_{\nu})
(\delta^{\nu}_{\sigma}-h^{\nu}_{\sigma}) u\indices{^{\sigma}_{;\beta}}h^{\beta}_{\mu} = u\indices{^{\nu}_{;\mu}}h^{\mu}_{\rho}u\indices{^{\rho}_{;\tau}} (\delta^{\tau}_{\nu}-h^{\tau}_{\nu}) \\ = 
(\ell\cdot u)^{-1} \ell_{\nu}u\indices{^{\nu}_{;\mu}} a^{\mu} - (\ell\cdot u)^{-2}(\ell\cdot a)^2 + (k\cdot v)^{-1} k_{\nu}u\indices{^{\nu}_{;\mu}} a^{\mu}_v - (k\cdot v)^{-2}(k\cdot a_v)^2 - 2(\ell\cdot u)^{-1}(k\cdot v)^{-1} (k\cdot a)(\ell \cdot a_v).
\end{multline}
Here we have introduced the notation
\begin{equation}
 a_v^{\mu} = u\indices{^{\mu}_{;\nu}}v^{\nu}.
\end{equation}
Meanwhile, using Eq.\ (\ref{Eq:NullProjection}) in $\dot{h}^{\beta}_{\alpha}u\indices{^{\alpha}_{;\beta}}$ results in
\begin{multline}
 \dot{h}^{\beta}_{\alpha}u\indices{^{\alpha}_{;\beta}} = (\ell\cdot u)^{-2}\dot{(\ell\cdot u)}(\ell\cdot a) + (k\cdot v)^{-2}\dot{(k\cdot v)}(k\cdot a_v) - (\ell\cdot u)^{-1}\left[(\dot{\ell}\cdot a)+\ell_{\alpha}u\indices{^{\alpha}_{;\beta}}a^{\beta}\right] - (k\cdot v)^{-1}\left[(\dot{k}\cdot a_v)+k_{\alpha}u\indices{^{\alpha}_{;\beta}}\dot{v}^{\beta}\right].
\end{multline}
Inserting these expressions into Eq.\ (\ref{Eq:ThetaDot1}) and collecting terms, we can write
\begin{multline} \label{Eq:ThetaDot2}
\dot{\Theta} =  
- B\indices{^{\mu}_{\nu}}B\indices{^{\nu}_{\mu}} 
- R\indices{^{\rho}_{\lambda\rho\tau}}u^{\lambda}u^{\tau} + a\indices{^{\rho}_{;\rho}} + 2(\ell\cdot u)^{-1}(k\cdot v)^{-1} (k\cdot a)(\ell \cdot a_v) \\
+ (\ell\cdot u)^{-2}\left[(\ell \cdot a)^2 + \dot{(\ell \cdot u)}(\ell\cdot a) -  (\ell \cdot u)\dot{(\ell\cdot a)} - 2(\ell\cdot u) \ell_{\alpha}u\indices{^{\alpha}_{;\beta}}a^{\beta} \right]\\
+ (k\cdot v)^{-2} \left[ (k\cdot a_v)^2 + \dot{(k\cdot v)}(k\cdot a_v) - (k\cdot v)\left( (\dot{k}\cdot a_v) + k_{\alpha}u\indices{^{\alpha}_{;\beta}}\dot{v}^{\beta} + k_{\alpha}u\indices{^{\alpha}_{;\beta}}a_v^{\beta} + k_{\alpha}v^{\beta}a\indices{^{\alpha}_{;\beta}} - k_{\alpha}R\indices{^{\alpha}_{\lambda\beta\sigma}}u^{\lambda}v^{\beta}u^{\sigma}\right) \right].
\end{multline}
Now use the identity
\begin{equation}
 \begin{aligned}
 k_{\alpha}u\indices{^{\alpha}_{;\beta}}\dot{v}^{\beta} & = k_{\alpha}u\indices{^{\alpha}_{;\beta}}v\indices{^{\beta}_{;\sigma}}u^{\sigma} = k_{\alpha}(u\indices{^{\alpha}_{;\beta}}v^{\beta})_{;\sigma}u^{\sigma} - k_{\alpha}u\indices{^{\alpha}_{;\beta\sigma}}v^{\beta}u^{\sigma}\\
 & = (k\cdot \dot{a}_v) - k_{\alpha}\left[u\indices{^{\alpha}_{;\sigma\beta}}u^{\sigma}v^{\beta} - R\indices{^{\alpha}_{\lambda\beta\sigma}}u^{\lambda}v^{\beta}u^{\sigma}\right] \\
 & = (k\cdot \dot{a}_v) - k_{\alpha}v^{\beta}a\indices{^{\alpha}_{;\beta}} + k_{\alpha}u\indices{^{\alpha}_{;\sigma}}a_v^{\sigma} + k_{\alpha}R\indices{^{\alpha}_{\lambda\beta\sigma}}u^{\lambda}v^{\beta}u^{\sigma}
 \end{aligned}
\end{equation}
to find
\begin{multline}\label{Eq:ThetaDot3}
\dot{\Theta} =  
- B\indices{^{\mu}_{\nu}}B\indices{^{\nu}_{\mu}} 
- R\indices{^{\rho}_{\lambda\rho\tau}}u^{\lambda}u^{\tau} + a\indices{^{\rho}_{;\rho}} + 2(\ell\cdot u)^{-1}(k\cdot v)^{-1} (k\cdot a)(\ell \cdot a_v) \\
+ (\ell\cdot u)^{-2}\left[(\ell \cdot a)^2 + \dot{(\ell \cdot u)}(\ell\cdot a) -  (\ell \cdot u)\dot{(\ell\cdot a)} - 2(\ell\cdot u) \ell_{\alpha}u\indices{^{\alpha}_{;\beta}}a^{\beta} \right]\\
+ (k\cdot v)^{-2} \left[ (k\cdot a_v)^2 + \dot{(k\cdot v)}(k\cdot a_v) - (k\cdot v)\dot{(k\cdot a_v)} - 2(k\cdot v)k_{\alpha}u\indices{^{\alpha}_{;\beta}}a_v^{\beta}\right].
\end{multline}
At this stage, we have managed to write the Raychaudhuri equation as the vacuum equation plus some additional terms.  
This is still not a very illuminating form for the Raychaudhuri equation since it depends on the unknown quantity $u\indices{^{\mu}_{;\nu}}$ and is not written in terms of the optical metric and its derivatives, as desired.
Even more importantly, it appears to depend on the arbitrarily chosen pair $(\bm{\ell},\bm{v})$, which is problematic for assigning any physical meaning to the equation.  It is instructive to look at some specific examples to examine the dependence on $(\bm{\ell},\bm{v})$.
\subsection{Nonaffine geodesics in vacuum}
In vacuum it is true that $k_{\alpha} = u_{\alpha}$.  Then $u_{\alpha}u\indices{^{\alpha}_{;\beta}} = \frac12 (u_{\alpha}u^{\alpha})_{;\beta} = 0$, from which it follows that $(k\cdot a) = 0$ and $(k\cdot a_v) =0$; thus
all such terms vanish from Eq.\ (\ref{Eq:ThetaDot3}).
To see that this expression reduces to the usual one in vacuum, consider the general case of nonaffinely parametrized geodesics such that $\bm{a} = \kappa \bm{u}$.  
One readily calculates
\begin{equation}
 \dot{\Theta} =  
 - B\indices{^{\mu}_{\nu}}B\indices{^{\nu}_{\mu}} 
 - R\indices{^{\rho}_{\lambda\rho\tau}}u^{\lambda}u^{\tau} + \kappa(u\indices{^{\rho}_{;\rho}} -\kappa).
\end{equation}
Since in this case
\begin{equation}
\Theta = h^{\alpha}_{\beta}u\indices{^{\beta}_{;\alpha}} = u\indices{^{\alpha}_{;\alpha}} - (\ell\cdot u)^{-1}(\ell\cdot a) = u\indices{^{\alpha}_{;\alpha}} - \kappa,
\end{equation}
we recover the usual expression in vacuum for nonaffinely parametrized geodesics
\begin{equation} \label{Eq:VacThetaDot}
 \dot{\Theta} =  
 - B\indices{^{\mu}_{\nu}}B\indices{^{\nu}_{\mu}} 
 - R\indices{^{\rho}_{\lambda\rho\tau}}u^{\lambda}u^{\tau} + \kappa\Theta.
\end{equation}

\subsection{Nongeodesic curves in media}
It turned out, somewhat miraculously, that in vacuum the pair $(\bm{\ell},\bm{v})$ completely dropped out of the final expression for $\dot{\Theta}$ in Eq.~(\ref{Eq:VacThetaDot}) and so may be chosen arbitrarily. 
This was a consequence of the fact that $\bm{k}$ in vacuum is orthogonal to both $\bm{a}$ and $\bm{a}_v$, and the geodesic nature of the curve that restricts $\bm{a}$ to the form $\bm{a} = \kappa \bm{u}$.
In general refracting media, we are faced with the fact that \textit{none} of these conditions hold.
In particular, since $\gamma_{\alpha\beta}u^{\alpha}u^{\beta} = 0$ implies
\begin{equation}
 \dot{\gamma}_{\alpha\beta}u^{\alpha}u^{\beta} + 2\gamma_{\alpha\beta}a^{\alpha}u^{\beta} = 0,
\end{equation}
we instead have that the most general form of $a^{\mu}$ admits
\begin{equation} \label{Eq:AccelerationGeneric}
 a^{\mu} = \kappa u^{\mu} + f\indices{^{\mu}_{\nu}}u^{\nu} - \frac12 \gamma^{\mu\alpha}\dot{\gamma}_{\alpha\nu}u^{\nu}
\end{equation}
where 
\begin{equation}
 f_{\alpha\nu} = \gamma_{\alpha\mu} f\indices{^{\mu}_{\nu}}
\end{equation}
is antisymmetric.  For congruences of timelike particles, $f\indices{^{\mu}_{\nu}}$ represents an external force on the particle.
It is unclear whether there exists a corresponding quantity for light in media, and in what follows we assume that this term does not contribute.

To compare with the vacuum result, consider just the parametrization artifact part of the acceleration $\kappa \bm{u}$.
First, note that now
\begin{equation}
 \kappa\Theta = \kappa h^{\alpha}_{\beta}u\indices{^{\beta}_{;\alpha}} = \kappa\left[u\indices{^{\alpha}_{;\alpha}} - \kappa - (k\cdot v)^{-1} (k\cdot a_v)\right].
\end{equation} 
All the terms proportional to $\bm{\ell}$ or $\bm{v}$ in Eq.\ (\ref{Eq:ThetaDot3}) still conspire to provide the $\kappa^2$ term in the expression for $\kappa\Theta$, but we are now left with 
\begin{equation}
 (k\cdot v)^{-2}\left[(k\cdot a_v)^2 + \dot{(k\cdot v)}(k\cdot a_v) - (k\cdot v)\left[\dot{(k\cdot a_v)} + 2k_{\alpha}u\indices{^{\alpha}_{;\beta}}a_v^{\beta} - \kappa (k\cdot a_v)\right] \right].
\end{equation} 
Thus, even for the relatively simple parametrization-dependent component of $\bm{a}$ we start seeing nontrivial dependence on the choice of $(\bm{\ell},\bm{v})$.
Of course, the parametrization may be chosen such that $\kappa=0$, but the other terms in $\bm{a}$ will contribute similarly, and the point is that the final answer for the expansion appears to depend on the choice of $(\bm{\ell},\bm{v})$.
This presents quite a big conceptual problem since at this stage the pair $(\bm{\ell},\bm{v})$ was chosen arbitrarily and so should carry no physical significance; thus, a result that depends on $\bm{v}$ is not well defined.

\subsection{Raychaudhuri equation for affinely parametrized congruences} \label{Sec:Raychaudhuri_last}

\begin{figure}[h]
 \scalebox{.65}{\includegraphics[]{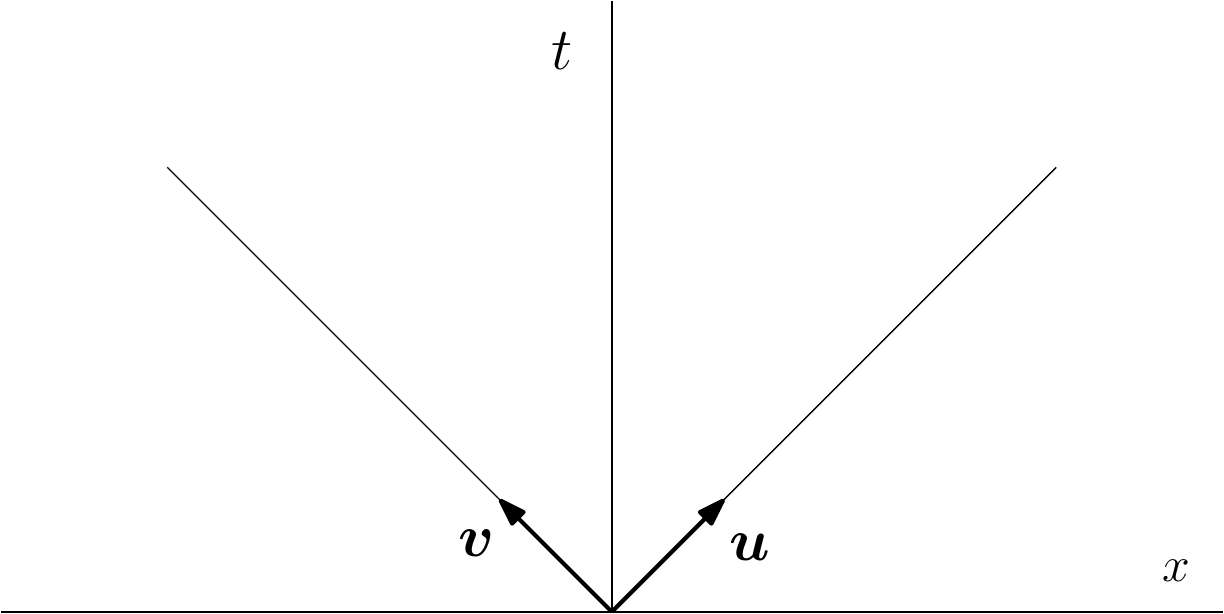}}
 \caption{A basis of vectors on the light cone spans the two-dimensional $tx$ subspace at each point.}
 \label{Fig:Lightcone}
\end{figure}

It is clear that we must look for a canonical choice of $(\bm{\ell},\bm{v})$ that carries some physical significance.
Ideally, the pair $(\bm{\ell},\bm{v})$ should be related by the same optical metric as the solution pair $(\bm{k},\bm{u})$, because $\gamma_{\alpha\beta}(\boldsymbol{\ell}) = \gamma_{\alpha\beta}(\bm{k})$ would significantly improve our ability to simplify both the projection operator $h^{\alpha}_{\beta}$ and ultimately the final answer.
Let us seek guidance from the simple case of a light ray propagating through vacuum in the $x$-direction.
In this simplified case the vector $\bm{u}$ is tangent to the solution curve $\tilde{\gamma}$ at the point $\tilde{\gamma}(\tau_0) = (t_0,x_0)$, as depicted in Fig.~\ref{Fig:Lightcone}.
If $\bm{t}$ is tangent to the coordinate line $t$ at this point, then 
\begin{equation} \label{Eq:vVacuum}
 \bm{v} = -\bm{u} - 2 \bm{g}(\bm{u},\bm{t}) \bm{t}
\end{equation}
is a future-directed null vector with spatial component along $-x$ and associated wave vector
\begin{equation} \label{Eq:ellVacuum}
 \bm{\ell} = \bm{v}^{\flat} = -\bm{k} - 2\bm{g}(\bm{u},\bm{t}) \bm{t}^{\flat}
\end{equation}
Thus, given $\bm{u}$ and the coordinate time $t$ of a local observer, we have constructed a second null vector $\bm{v}$ such that $\bm{u}$ and $\bm{v}$ span the light cone at $\tilde{\gamma}(\tau_0)$.

It is important to note that $(\bm{\ell},\bm{v})$ is not necessarily a solution in the sense that $\bm{v}$ is tangent to a solution curve with $\bm{\ell} = \ed S'$ for phase function $S'$; i.e., we have not shown whether, or under what conditions, $\bm{\ell}$ is exact.
To be specific, the time coordinate gives us a curve in a local chart $t:M\to \mathbb{R}$.
The exterior derivative of this $\mathbb{R}$-valued function on $M$ is $\ed t$.
If the space-time is static, then $(\ed t)^{\sharp}=\frac{\partial}{\partial t} = \bm{t}$, which is indeed the expected tangent vector to the coordinate $t$. 
Therefore, if the space-time is static, we have
\begin{equation}
 \bm{\ell} = -\ed S - 2\bm{g}(\bm{u},\bm{t}) \ed t = \ed S'
\end{equation}
where the last equality holds because in vacuum we can always scale the null vector $\bm{u}$ such that $\ed(\bm{g}(\bm{u},\bm{t}))=0$. 
Thus if we consider $\bm{u}$ as tangent to the ``outgoing'' ray, then in \textit{static} space-times, the complimentary null vector $\bm{v}$ is indeed tangent to the ``ingoing'' ray.
The condition of static space-times indicates a sense of reciprocity: if a wave front of the outgoing ray is retroreflected, it will follow the same spatial path through the manifold, passing back through the point $x_0$ with tangent $\bm{v}$.

Taking this reasoning as a starting point, let $t^{\mu}$ be a vector field parallel to the world lines of a family of inertial observers who measure the medium parameters, but normalized with $\gamma_{\alpha\beta}(k)t^{\alpha}t^{\beta} = -1$.
Since the optical metric is conformally invariant, it can always be arranged that $\gamma_{\alpha\beta}t^{\alpha}t^{\beta} = g_{\alpha\beta}t^{\alpha}t^{\beta}=-1$, so this normalization is indeed the natural one for the inertial observers.
Now let the pair $(\bm{\ell},\bm{v})$ be defined similarly to Eqs.~(\ref{Eq:vVacuum}) and (\ref{Eq:ellVacuum}), but where the optical metric replaces the space-time metric
 
\begin{align} 
 v^{\mu} &  = -u^{\mu} - 2(\bm{k}\cdot\bm{t})t^{\mu} \label{Eq:v}, \\
 \ell_{\nu} & = \gamma_{\nu\mu}(k) v^{\mu} = -k_{\nu} - 2(\bm{k}\cdot\bm{t})\gamma_{\nu\mu}t^{\mu} \label{Eq:ell}.
\end{align}
Here we have written $(\bm{k}\cdot \bm{t}) = \gamma_{\mu\nu}u^{\mu}t^{\nu}$ as a shorthand reminder that the product is with respect to the optical metric rather than the space-time metric, as in Eq.~(\ref{Eq:vVacuum}).
It is straightforward to show that $(\bm{\ell}\cdot \bm{u}) = (\bm{k}\cdot \bm{v}) = -2(\bm{k}\cdot \bm{t})^2$.

Although $(\bm{\ell},\bm{v})$ completes the null basis of the light cone as in the vacuum case, it does not necessarily correspond to an ingoing solution curve $\gamma'$.  For this to happen, we would further require
\begin{equation}
 \boldsymbol{\gamma}(\bm{t},\cdot) = \bm{t}^{\flat} = \ed t
\end{equation}
which requires not only static space-times but reciprocal media.
Since all we need is a complete null basis at each point visited by the reference curve of the congruence, the construction of $(\bm{\ell},\bm{v})$ by Eqs.~(\ref{Eq:v}) and (\ref{Eq:ell}) is sufficient to trace the congruence $\tilde{\gamma}$.
In vacuum, most treatments scale $(\bm{\ell}\cdot \bm{u}) = (\bm{k}\cdot \bm{v}) = -1$.  
In media, however, such a scaling would require further assumptions on the medium.  The quantity $(\bm{k}\cdot \bm{t})$ has a physical interpretation as the frequency measured by the observer.
For interesting applications, for example in transformation optics, it is desirable to allow for frequency-modulating media \cite{Cummer2011jo}. 
For this reason we do not require $(\bm{k}\cdot \bm{t})$ to be constant.

Lastly, we assume that the congruence is affinely parametrized and that the only contribution to $\bm{a}$ from Eq.\ (\ref{Eq:AccelerationGeneric}) is
\begin{equation}
 a^{\alpha} = -\frac12 \gamma^{\mu\alpha}\dot{\gamma}_{\mu\nu}u ^{\nu}.
\end{equation}

Recall that our task is to calculate
\begin{equation}
\dot{\Theta} = 
- B\indices{^{\mu}_{\nu}}B\indices{^{\nu}_{\mu}} 
- h_{\alpha}^{\beta}R\indices{^{\alpha}_{\lambda\beta\tau}}u^{\lambda}u^{\tau} + h_{\alpha}^{\beta}a\indices{^{\alpha}_{;\beta}} + \dot{h}^{\beta}_{\rho}u\indices{^{\rho}_{;\beta}} -h^{\mu}_{\rho}u\indices{^{\rho}_{;\tau}} (\delta^{\tau}_{\nu}-h^{\tau}_{\nu})
(\delta^{\nu}_{\sigma}-h^{\nu}_{\sigma}) u\indices{^{\sigma}_{;\beta}}h^{\beta}_{\mu} 
\end{equation}
With the physically meaningful construction of $(\bm{\ell},\bm{u})$ given by Eqs.~(\ref{Eq:v}) and (\ref{Eq:ell}) now at our disposal, we find it most straightforward to return to the beginning and calculate each unknown term in turn.
The projection operator becomes
\begin{equation} 
 h_{\alpha}^{\beta}= \delta_{\alpha}^{\beta} - (\bm{k}\cdot\bm{t})^{-2}\gamma_{\mu\alpha}(u^{\mu}u^{\beta}+(\bm{k}\cdot\bm{t})(t^{\mu}u^{\beta} + t^{\beta}u^{\mu})).
\end{equation}
The details of the lengthy calculation are presented in the Appendix. The result is
\begin{multline}
 \dot{\Theta} = -B\indices{^{\mu}_{\nu}}B\indices{^{\nu}_{\mu}} - \frac12 R\indices{_{\beta\tau\alpha}^{\nu}} (g^{\alpha\beta}g_{\nu\lambda}+\gamma^{\alpha\beta}\gamma_{\nu\lambda})u^{\tau}u^{\lambda}
 -\frac12 \gamma^{\alpha\sigma}\gamma_{\sigma\rho;\tau} (B\indices{^{\rho}_{\alpha}}u^{\tau}+B\indices{^{\tau}_{\alpha}}u^{\rho}) \\
 -\frac12 h^{\beta}_{\alpha}\left[\gamma^{\alpha\sigma}\gamma_{\sigma\rho;\beta\tau}u^{\rho}u^{\tau} + \gamma\indices{^{\alpha\sigma}_{;\beta}}\dot{\gamma}_{\sigma\rho}u^{\rho} + \frac12 \gamma^{\sigma\omega}\dot{\gamma}_{\beta\sigma}\gamma^{\alpha\lambda}\dot{\gamma}_{\lambda\mu}(\delta^{\mu}_{\omega}-h^{\mu}_{\omega}) \right].
\end{multline}

It is easy to see that this expression reduces to the usual one in the vacuum limit $\gamma_{\alpha\beta}\to g_{\alpha\beta}$ since the covariant derivative $g_{\alpha\beta;\mu} =0$, but there are some other interesting features to note.
The first is that although we might have expected to be able to identify a Riemann-like tensor term built out of second derivatives of the optical metric we find this is not the case.  
A single second derivative term appears, but this is not enough to identify a Riemann-like tensor from which we can define an effective curvature of the medium.
Also of note is that this expression still depends on $\bm{t}$ through the dependence on the projection operator $h^{\beta}_{\alpha}$.
We earlier found that the arbitrarily chosen null vector $\bm{v}$ appeared in $\dot{\Theta}$, which forced us to seek a canonical choice of $\bm{v}$ that reflected some physical meaning, which we did via the world lines of a preferred set of observers -- those who make measurements of the material parameters and the congruence. 
So, we should not be too surprised by the continued presence of $\bm{t}$ in the final expression, but one might have harbored some hope that the $\bm{t}$-dependence would drop out by some fortuitous cancellation.
Another interesting and unexpected feature is the interaction of the optical metric and the background metric through the Riemann tensor.
In the vacuum limit, these terms combine to give us the usual Riemann tensor term, but here we see that this contribution is actually split between the background metric and the optical metric.

In vacuum and for nonrotating congruences, one finds that $\dot{\Theta}$ is driven by the Riemann tensor term, and the positivity of space-time curvature forces a focusing of the congruence. 
In flat space-time, the Riemann tensor vanishes, and $\dot{\Theta}$ is primarily driven by first and second derivatives of the optical metric.
These derivative terms can contribute with either sign, leading to a corresponding focusing or defocusing of the congruence.

Further insight may be gained through more detailed knowledge of the optical metric $\gamma^{\alpha\beta}$.
A better understanding of how the individual dielectric parameters of permeability, permittivity, and magnetoelectric couplings contribute to the optical metric would show how they each contribute to the focusing of a beam.
The optical metric must contain contributions from both the background metric and the medium parameters, but an expression for $\gamma^{\alpha\beta}$ in terms of $g^{\alpha\beta}$ and $\chi\indices{_{\alpha\beta}^{\mu\nu}}$ is not currently known.
Since the medium parameters are also measured by the inertial observers $\bm{t}$, it is possible that the unknown expression for $\gamma^{\alpha\beta}$ could be formulated in such a way that $\bm{t}$ also appears.

\section{Conclusions} \label{Sec:Conclusions}


In the past decade, transformation optics has renewed interest in the longstanding recognition of the similarity of the refractive properties of dielectric media to those of curved space-times.
This has inspired a program that seeks to apply a range of tools and techniques from general relativity to the analysis of electrodynamics in dielectric media.
In particular, several authors are working on covariant formulations of electrodynamics within dielectric media.
In our approach, we recognize that any real medium must reside within a background space-time, and that the structure imposed by the medium is supplemental to that of the background space-time.
It is desirable, therefore, to have a framework for electrodynamics in media that explicitly and separately accounts for the medium and the background.

These considerations motivate us to study the detailed behavior of light propagation through linear dielectric media residing within a curved background space-time.  
We have derived the covariant kinematics for congruences of light in dielectric media and used this to formulate a generalized Raychaudhuri equation in media.
The covariant kinematics allow us to go beyond ray trajectories to analyze physically meaningful and measurable characteristics of beam propagation like the expansion, shear, and rotation of the cross section of a beam.

Since the behavior of congruences, governed by the Raychaudhuri equation, serves as a probe of curvature, we supposed that the derivation of a generalized Raychaudhuri equation in media should allow the identification of a Riemann-like tensor for the medium.
Our expectation was that we would be able to identify such a tensor term in the generalized Raychaudhuri equation, comprised of combinations of second derivatives of the optical metric, and which does not vanish for media within Minkowski space-time.

Our derivation showed that the Raychaudhuri equation for media residing in flat space-time is dominated by first and second derivatives of the optical metric but that these quantities do not appear to have any special corresponding physical interpretation in general relativity.
We are thus unable to identify a Riemann-like tensor describing the curvature of linear dielectric media.
One interesting feature of the generalized Raychaudhuri equation is a dependence on the auxiliary pair $(\bm{v},\bm{\ell})$.  
In the vacuum derivation the auxiliary vector drops out of the final result and so may be chosen arbitrarily, but its continued presence in the generalized equation forced us to seek a canonical construction of $(\bm{v},\bm{\ell})$, which was possible through the choice of a timelike vector field $\bm{t}$.
We argue that this choice is canonical and corresponds to the world lines of a family of timelike observers who are performing the measurements.

The techniques and results presented here can be used for more accurate congruence tracing through dielectric media, which should be useful for studying the behavior of optical devices, particularly those devised through transformation optics.
For example, we have used these results to study the expansion of congruences for the case of a dielectric analog space-time where the analog is taken to reside in a flat background space-time \cite{Fathi2016prd}.

Although this work was primarily motivated by questions in transformation optics and dielectric analog space-times, it would also be possible and interesting to apply these results to ray or congruence tracing in astrophysical settings.
For example radio waves propagating through dust clouds or accretion disks near massive compact objects could experience some dielectric refraction in addition to the effects of space-time curvature.
Such refractive effects could have implications for cosmological surveys based on gravitational lensing.

Lastly, we have so far only considered the first order behavior of the transverse vector $\bar{\xi}$, as can be seen in Eq.~(\ref{Eq:TaylorExp}).
By extending this work to the second order behavior it would be possible to generalize the geodesic deviation equation to light in media.
Such a generalized deviation equation would provide the starting point for even more accurate ray tracing through dielectric media.
We leave this derivation as the subject of future study.

\appendix

\section{Detailed calculation of Raychaudhuri equation in media}
In this Appendix we present a fully detailed account of the calculations for the Raychaudhuri equation presented in Sec.\ \ref{Sec:Raychaudhuri_last}.  The Raychaudhuri equation is
\begin{equation} \label{Eq:ThetaDot}
\dot{\Theta} = 
- B\indices{^{\mu}_{\nu}}B\indices{^{\nu}_{\mu}} 
- h_{\alpha}^{\beta}R\indices{^{\alpha}_{\lambda\beta\tau}}u^{\lambda}u^{\tau} + h_{\alpha}^{\beta}a\indices{^{\alpha}_{;\beta}} + \dot{h}^{\beta}_{\rho}u\indices{^{\rho}_{;\beta}} -h^{\mu}_{\rho}u\indices{^{\rho}_{;\tau}} (\delta^{\tau}_{\nu}-h^{\tau}_{\nu})
(\delta^{\nu}_{\sigma}-h^{\nu}_{\sigma}) u\indices{^{\sigma}_{;\beta}}h^{\beta}_{\mu} .
\end{equation}
With the assumptions made in Sec.\ \ref{Sec:Raychaudhuri_last}, the projection operator becomes
\begin{equation} \label{Eq:h}
 h_{\alpha}^{\beta}= \delta_{\alpha}^{\beta} - (\bm{k}\cdot\bm{t})^{-2}\gamma_{\mu\alpha}(u^{\mu}u^{\beta}+(\bm{k}\cdot\bm{t})(t^{\mu}u^{\beta} + t^{\beta}u^{\mu})).
\end{equation}
We will require $\dot{h}^{\beta}_{\alpha}$:
\begin{equation} \label{Eq:hdot}
 \begin{aligned}
   \dot{h}_{\alpha}^{\beta} 
   & = 2\dot{(\bm{k}\cdot\bm{t})}(\bm{k}\cdot\bm{t})^{-3}\gamma_{\mu\alpha}\left[u^{\mu}u^{\beta}+(\bm{k}\cdot\bm{t})(t^{\mu}u^{\beta} + t^{\beta}u^{\mu})\right] - \dot{(\bm{k}\cdot\bm{t})}(\bm{k}\cdot\bm{t})^{-3}\gamma_{\mu\alpha}\left[(\bm{k}\cdot\bm{t})(t^{\mu}u^{\beta} + t^{\beta}u^{\mu})\right] \\
     & \qquad -(\bm{k}\cdot\bm{t})^{-2}\dot{\gamma}_{\mu\alpha}\left[u^{\mu}u^{\beta}+(\bm{k}\cdot\bm{t})(t^{\mu}u^{\beta}+t^{\beta}u^{\mu})\right] - (\bm{k}\cdot\bm{t})^{-2}\gamma_{\mu\alpha}\left[a^{\mu}(u^{\beta}+(\bm{k}\cdot\bm{t})t^{\beta}) + a^{\beta}(u^{\mu}+(\bm{k}\cdot\bm{t})t^{\mu}) \right] \\ 
     & \qquad -(\bm{k}\cdot\bm{t})^{-1}\gamma_{\mu\alpha}[u^{\beta} \dot{t}^{\mu}+ u^{\mu}\dot{t}^{\beta}]\\
   & = \dot{(\bm{k}\cdot\bm{t})}(\bm{k}\cdot\bm{t})^{-3}\gamma_{\mu\alpha}[2u^{\mu}u^{\beta}+(\bm{k}\cdot\bm{t})(t^{\mu}u^{\beta} + t^{\beta}u^{\mu})] -(\bm{k}\cdot\bm{t})^{-2}\dot{\gamma}_{\mu\alpha}\left[u^{\mu}(u^{\beta}+(\bm{k}\cdot\bm{t})t^{\beta}) + (\bm{k}\cdot\bm{t})t^{\mu}u^{\beta}\right] \\ 
     & \qquad + \frac12 (\bm{k}\cdot\bm{t})^{-2}\gamma_{\mu\alpha}\left[\gamma^{\mu\sigma}\dot{\gamma}_{\sigma\rho}u^{\rho}(u^{\beta}+(\bm{k}\cdot\bm{t})t^{\beta}) + \gamma^{\beta\sigma}\dot{\gamma}_{\sigma\rho}u^{\rho}(u^{\mu}+(\bm{k}\cdot\bm{t})t^{\mu}) \right] -(\bm{k}\cdot\bm{t})^{-1}\gamma_{\mu\alpha}[u^{\beta} \dot{t}^{\mu}+ u^{\mu}\dot{t}^{\beta}]\\
   & = \dot{(\bm{k}\cdot\bm{t})}(\bm{k}\cdot\bm{t})^{-3}\gamma_{\mu\alpha}[2u^{\mu}u^{\beta}+(\bm{k}\cdot\bm{t})(t^{\mu}u^{\beta} + t^{\beta}u^{\mu})] -\frac12 (\bm{k}\cdot\bm{t})^{-2} \dot{\gamma}_{\mu\alpha}u^{\mu}[u^{\beta}+(\bm{k}\cdot\bm{t})t^{\beta}]  \\ 
     & \qquad + \frac12 (\bm{k}\cdot\bm{t})^{-2} \gamma_{\mu\alpha} \gamma^{\beta\sigma}\dot{\gamma}_{\sigma\rho} u^{\rho}[u^{\mu}+(\bm{k}\cdot\bm{t})t^{\mu}] 
     -(\bm{k}\cdot\bm{t})^{-1}[\gamma_{\mu\alpha}u^{\mu} \dot{t}^{\beta} + u^{\beta}\dot{(\gamma_{\alpha\mu}t^{\mu})}].
 \end{aligned}
\end{equation}

We calculate each unknown term of Eq.\ (\ref{Eq:ThetaDot}) in the following subsections.

\subsection{$h_{\alpha}^{\beta}R\indices{^{\alpha}_{\lambda\beta\tau}}u^{\lambda}u^{\tau}$}

\begin{equation} \label{Eq:Term1}
h_{\alpha}^{\beta}R\indices{^{\alpha}_{\lambda\beta\tau}}u^{\lambda}u^{\tau} = R\indices{^{\alpha}_{\lambda\alpha\tau}}u^{\lambda}u^{\tau} - (\bm{k}\cdot\bm{t})^{-1}\gamma_{\mu\alpha}u^{\mu}R\indices{^{\alpha}_{\lambda\beta\tau}}u^{\lambda}t^{\beta}u^{\tau}.
\end{equation}

\subsection{$h_{\alpha}^{\beta}a\indices{^{\alpha}_{;\beta}}$}
We assume that the only contribution to the acceleration comes from the medium term
\begin{equation}
 a^{\mu} = -\frac12 \gamma^{\mu\sigma}\dot{\gamma}_{\sigma\nu}u^{\nu}.
\end{equation}
First take the derivative
\begin{equation}
 a\indices{^{\alpha}_{;\beta}} = -\frac12 \left[ \left( \gamma\indices{^{\alpha\sigma}_{;\beta}}\gamma_{\sigma\rho;\tau} + \gamma^{\alpha\sigma}\gamma_{\sigma\rho;\tau\beta} \right)u^{\rho}u^{\tau} + \gamma^{\alpha\sigma}\gamma_{\sigma\rho;\tau}\left(u\indices{^{\rho}_{;\beta}}u^{\tau} + u^{\rho}u\indices{^{\tau}_{;\beta}} \right)  \right].
\end{equation}
Next, operate with the projection operator
\begin{equation}
h_{\alpha}^{\beta}a\indices{^{\alpha}_{;\beta}} = -\frac12 h_{\alpha}^{\beta}\left( \gamma\indices{^{\alpha\sigma}_{;\beta}}\gamma_{\sigma\rho;\tau} + \gamma^{\alpha\sigma}\gamma_{\sigma\rho;\tau\beta} \right)u^{\rho}u^{\tau} - \frac12 \gamma^{\alpha\sigma}\gamma_{\sigma\rho;\tau}\left(u\indices{^{\rho}_{;\beta}}h_{\alpha}^{\beta}u^{\tau} + u^{\rho}u\indices{^{\tau}_{;\beta}}h^{\beta}_{\alpha} \right).
\end{equation}
We can rewrite this with some clever addition and subtraction:
\begin{multline}
h_{\alpha}^{\beta}a\indices{^{\alpha}_{;\beta}} = -\frac12 h_{\alpha}^{\beta}\left( \gamma\indices{^{\alpha\sigma}_{;\beta}}\gamma_{\sigma\rho;\tau} + \gamma^{\alpha\sigma}\gamma_{\sigma\rho;\tau\beta} \right)u^{\rho}u^{\tau} \\
- \frac12 \gamma^{\alpha\sigma}\gamma_{\sigma\rho;\tau} \left[h_{\lambda}^{\rho}u\indices{^{\lambda}_{;\beta}}h^{\beta}_{\alpha} + (\delta_{\lambda}^{\rho}-h_{\lambda}^{\rho})\dot{h}_{\alpha}^{\lambda} + (\delta_{\beta}^{\rho}-h_{\beta}^{\rho})(u\indices{^{\beta}_{;\lambda}}h^{\lambda}_{\alpha} - \dot{h}_{\alpha}^{\beta}) \right]u^{\tau} \\
- \frac12 \gamma^{\alpha\sigma}\gamma_{\sigma\rho;\tau} \left[h_{\lambda}^{\tau}u\indices{^{\lambda}_{;\beta}}h^{\beta}_{\alpha} + (\delta_{\lambda}^{\tau}-h_{\lambda}^{\tau})\dot{h}_{\alpha}^{\lambda} + (\delta_{\beta}^{\tau}-h_{\beta}^{\tau})(u\indices{^{\beta}_{;\lambda}}h^{\lambda}_{\alpha} - \dot{h}_{\alpha}^{\beta}) \right]u^{\rho}.
\end{multline}
By Eq.~(\ref{Eq:B}), this now becomes
\begin{multline}
h_{\alpha}^{\beta}a\indices{^{\alpha}_{;\beta}} = -\frac12 h_{\alpha}^{\beta}\left( \gamma\indices{^{\alpha\sigma}_{;\beta}}\gamma_{\sigma\rho;\tau} + \gamma^{\alpha\sigma}\gamma_{\sigma\rho;\tau\beta} \right)u^{\rho}u^{\tau} - \frac12 \gamma^{\alpha\sigma}\gamma_{\sigma\rho;\tau} \left( B\indices{^{\rho}_{\alpha}}u^{\tau} + B\indices{^{\tau}_{\alpha}}u^{\rho}\right)  \\
- \frac12 \gamma^{\alpha\sigma}\gamma_{\sigma\rho;\tau} \left[ u^{\tau} (\delta_{\beta}^{\rho}-h_{\beta}^{\rho}) (u\indices{^{\beta}_{;\lambda}}h^{\lambda}_{\alpha} - \dot{h}_{\alpha}^{\beta}) + u^{\rho}  (\delta_{\beta}^{\tau}-h_{\beta}^{\tau}) (u\indices{^{\beta}_{;\lambda}}h^{\lambda}_{\alpha} - \dot{h}_{\alpha}^{\beta}) \right]. 
\end{multline}
Consider the term $(\delta_{\beta}^{\rho}-h_{\beta}^{\rho})(u\indices{^{\beta}_{;\lambda}}h^{\lambda}_{\alpha} - \dot{h}_{\alpha}^{\beta})$.   We have already calculated $\dot{h}^{\beta}_{\alpha}$ above, so first calculate
\begin{equation}
 \begin{aligned}
 u\indices{^{\beta}_{;\lambda}}h^{\lambda}_{\alpha} 
 & =  u\indices{^{\beta}_{;\alpha}} - (\bm{k}\cdot\bm{t})^{-2} \gamma_{\mu\alpha}a^{\beta}[u^{\mu}+(\bm{k}\cdot\bm{t})t^{\mu}] - (\bm{k}\cdot\bm{t})^{-1} \gamma_{\mu\alpha}u^{\mu}(t^{\lambda}u\indices{^{\beta}_{;\lambda}})\\
 & =  u\indices{^{\beta}_{;\alpha}} + \frac12 (\bm{k}\cdot\bm{t})^{-2} \gamma_{\mu\alpha}\gamma^{\beta\sigma}\dot{\gamma}_{\sigma\rho}u^{\rho}[u^{\mu}+(\bm{k}\cdot\bm{t})t^{\mu}] - (\bm{k}\cdot\bm{t})^{-1} \gamma_{\mu\alpha}u^{\mu}(t^{\lambda}u\indices{^{\beta}_{;\lambda}}),
 \end{aligned}
\end{equation}
then combine with Eq.~(\ref{Eq:hdot}) to get
\begin{multline}
 u\indices{^{\beta}_{;\lambda}}h^{\lambda}_{\alpha} - \dot{h}_{\alpha}^{\beta} 
  = u\indices{^{\beta}_{;\alpha}} + \frac12 (\bm{k}\cdot\bm{t})^{-2} \dot{\gamma}_{\alpha\mu}u^{\mu}[u^{\beta} + (\bm{k}\cdot\bm{t})t^{\beta}]
 - (\bm{k}\cdot\bm{t})^{-1}[\gamma_{\alpha\mu}u^{\mu}(t^{\lambda}u\indices{^{\beta}_{;\lambda}}-u^{\lambda}t\indices{^{\beta}_{;\lambda}}) - u^{\beta}\dot{(\gamma_{\alpha\mu}t^{\mu})}]  \\
  - \dot{(\bm{k}\cdot\bm{t})} (\bm{k}\cdot\bm{t})^{-3}\gamma_{\alpha\mu}[u^{\beta}(u^{\mu}+(\bm{k}\cdot\bm{t})t^{\mu}) + u^{\mu}(u^{\beta}+(\bm{k}\cdot\bm{t}) t^{\beta})].
\end{multline}
To get $(\delta_{\beta}^{\rho}-h_{\beta}^{\rho})(u\indices{^{\beta}_{;\lambda}}h^{\lambda}_{\alpha} - \dot{h}_{\alpha}^{\beta})$, first note that $h^{\alpha}_{\rho}$ annihilates both $u^{\rho}$ and $t^{\rho}$, so  $(\delta_{\beta}^{\rho}-h_{\beta}^{\rho})u^{\beta} = u^{\rho}$, and similarly for $t^{\beta}$. Thus
\begin{equation}
 \begin{aligned}
  (\delta_{\beta}^{\rho}-h_{\beta}^{\rho})(u\indices{^{\beta}_{;\lambda}}h^{\lambda}_{\alpha} - \dot{h}_{\alpha}^{\beta})
  & =  
 (\delta_{\beta}^{\rho}-h_{\beta}^{\rho})u\indices{^{\beta}_{;\alpha}} + \frac12 (\bm{k}\cdot\bm{t})^{-2} \dot{\gamma}_{\alpha\mu}u^{\mu}[u^{\rho} + (\bm{k}\cdot\bm{t})t^{\rho}] \\
    & \qquad - (\bm{k}\cdot\bm{t})^{-1} [\gamma_{\alpha\mu}u^{\mu} (\delta_{\beta}^{\rho}-h_{\beta}^{\rho}) (t^{\lambda} u\indices{^{\beta}_{;\lambda}} -u^{\lambda} t\indices{^{\beta}_{;\lambda}}) - u^{\rho} \dot{(\gamma_{\alpha\mu} t^{\mu})}]  \\
    & \qquad - \dot{(\bm{k}\cdot\bm{t})} (\bm{k}\cdot\bm{t})^{-3}\gamma_{\alpha\mu}[u^{\rho}(u^{\mu}+(\bm{k}\cdot\bm{t})t^{\mu}) + u^{\mu}(u^{\rho}+(\bm{k}\cdot\bm{t}) t^{\rho})].
 \end{aligned}
\end{equation}
So there are only two terms left to deal with.  First,
\begin{equation}
 \begin{aligned}
 (\delta_{\beta}^{\rho}-h_{\beta}^{\rho})u\indices{^{\beta}_{;\alpha}} & = (\bm{k}\cdot\bm{t})^{-2}(\gamma_{\beta\mu}u^{\mu}u\indices{^{\beta}_{;\alpha}})(u^{\rho}+ (\bm{k}\cdot\bm{t})t^{\rho}) + (\bm{k}\cdot\bm{t})^{-1} u^{\rho}(\gamma_{\beta\mu}t^{\mu}u\indices{^{\beta}_{;\alpha}}) \\
 & = -\frac12 (\bm{k}\cdot\bm{t})^{-2}\dot{\gamma}_{\alpha\mu}u^{\mu}(u^{\rho}+ (\bm{k}\cdot\bm{t})t^{\rho}) + (\bm{k}\cdot\bm{t})^{-1} u^{\rho}(\gamma_{\beta\mu}t^{\mu}u\indices{^{\beta}_{;\alpha}})
 \end{aligned}
\end{equation}
where going from the first to the second line used a very useful identity that will be used several times again: 
\begin{equation} \label{Eqs:u-t-identities}
  \gamma_{\nu\lambda}u^{\nu}u\indices{^{\lambda}_{;\alpha}} = -(\gamma_{\nu\lambda}u^{\nu})_{;\alpha}u^{\lambda} = -k_{\lambda;\alpha}u^{\lambda} = -k_{\alpha;\lambda}u^{\lambda} = -\dot{k}_{\alpha} = -\frac12 \dot{\gamma}_{\alpha\mu}u^{\mu}.
\end{equation}
Secondly,
\begin{equation}
 \begin{aligned}
   (\delta_{\beta}^{\rho}-h_{\beta}^{\rho}) (t^{\lambda} u\indices{^{\beta}_{;\lambda}} -u^{\lambda} t\indices{^{\beta}_{;\lambda}}) 
   & = (\bm{k}\cdot\bm{t})^{-2} \left[ (\gamma_{\beta\mu}u^{\mu}u\indices{^{\beta}_{;\lambda}}) t^{\lambda} - (\gamma_{\beta\mu}u^{\mu}t\indices{^{\beta}_{;\lambda}})u^{\lambda}  \right](u^{\rho}+(\bm{k}\cdot\bm{t}) t^{\rho}) \\
     & \qquad + (\bm{k}\cdot\bm{t})^{-1}\gamma_{\beta\mu}t^{\mu}u^{\rho}(t^{\lambda}u\indices{^{\beta}_{;\lambda}} - u^{\lambda}t\indices{^{\beta}_{;\lambda}}) \\
   & = (\bm{k}\cdot\bm{t})^{-2} \left[ -\frac12 \dot{\gamma}_{\mu\lambda} u^{\mu} t^{\lambda} - \gamma_{\beta\mu}u^{\mu}\dot{t}^{\beta}  \right](u^{\rho}+(\bm{k}\cdot\bm{t}) t^{\rho}) 
       + (\bm{k}\cdot\bm{t})^{-1}\gamma_{\beta\mu}t^{\mu}u^{\rho}(t^{\lambda}u\indices{^{\beta}_{;\lambda}} - u^{\lambda}t\indices{^{\beta}_{;\lambda}}) \\   
   & = -(\bm{k}\cdot\bm{t})^{-2} \dot{(\bm{k}\cdot\bm{t})}(u^{\rho}+(\bm{k}\cdot\bm{t}) t^{\rho}) 
          + (\bm{k}\cdot\bm{t})^{-1}\gamma_{\beta\mu}t^{\mu}u^{\rho}(t^{\lambda}u\indices{^{\beta}_{;\lambda}} - u^{\lambda}t\indices{^{\beta}_{;\lambda}})
 \end{aligned}
\end{equation}
so that 
\begin{multline}
(\delta_{\beta}^{\rho}-h_{\beta}^{\rho})(u\indices{^{\beta}_{;\lambda}}h^{\lambda}_{\alpha} - \dot{h}_{\alpha}^{\beta}) = (\bm{k}\cdot\bm{t})^{-1}u^{\rho}[(\gamma_{\beta\mu}t^{\mu}u\indices{^{\beta}_{;\alpha}}) + \dot{(\gamma_{\alpha\mu}t^{\mu})}] \\ - (\bm{k}\cdot\bm{t})^{-2}u^{\rho}\gamma_{\alpha\mu}\gamma_{\beta\nu}u^{\mu}t^{\nu}[t^{\lambda} u\indices{^{\beta}_{;\lambda}} -u^{\lambda} t\indices{^{\beta}_{;\lambda}}] - (\bm{k}\cdot\bm{t})^{-3} \dot{(\bm{k}\cdot\bm{t})}u^{\rho}\gamma_{\alpha\mu}[u^{\mu}+(\bm{k}\cdot\bm{t}) t^{\mu}].
\end{multline}
Returning to $h_{\alpha}^{\beta}a\indices{^{\alpha}_{;\beta}}$ we can now find
\begin{multline}
  h_{\alpha}^{\beta}a\indices{^{\alpha}_{;\beta}}  = -\frac12h_{\alpha}^{\beta}(\gamma\indices{^{\alpha\sigma}_{;\beta}}\gamma_{\sigma\rho;\tau} + \gamma^{\alpha\sigma}\gamma_{\sigma\rho;\tau\beta})u^{\rho}u^{\tau} - \frac12 \gamma^{\alpha\sigma}\gamma_{\sigma\rho;\tau}(B\indices{^{\rho}_{\alpha}}u^{\tau} + B\indices{^{\tau}_{\alpha}}u^{\rho} ) - (\bm{k}\cdot\bm{t})^{-1}\gamma^{\alpha\sigma}\dot{\gamma}_{\sigma\rho}u^{\rho}\left[ (\gamma_{\lambda\mu}t^{\mu}u\indices{^{\lambda}_{;\alpha}}) + \dot{(\gamma_{\alpha\mu}t^{\mu})} \right]\\
  + (\bm{k}\cdot\bm{t})^{-2}(\dot{\gamma}_{\sigma\rho}u^{\sigma}u^{\rho}) \gamma_{\beta\nu}t^{\nu}[t^{\lambda} u\indices{^{\beta}_{;\lambda}} -u^{\lambda} t\indices{^{\beta}_{;\lambda}}] 
  + (\bm{k}\cdot\bm{t})^{-3} \dot{(\bm{k}\cdot\bm{t})}\dot{\gamma}_{\alpha\mu}u^{\alpha}[u^{\mu}+(\bm{k}\cdot\bm{t}) t^{\mu}].
\end{multline}
We commute the covariant derivatives on 
\begin{equation}
  \gamma_{\sigma\rho;\tau\beta} = \gamma_{\sigma\rho;\beta\tau} - \gamma_{\nu\rho}R\indices{^{\nu}_{\sigma\beta\tau}} - \gamma_{\nu\sigma}R\indices{^{\nu}_{\rho\beta\tau}} 
\end{equation}
and expand the projection operator on
\begin{equation}
 h^{\beta}_{\alpha}\gamma^{\alpha\sigma}\gamma_{\sigma\rho;\tau\beta}u^{\rho}u^{\tau} = h^{\beta}_{\alpha}\gamma^{\alpha\sigma}\gamma_{\sigma\rho;\beta\tau}u^{\rho}u^{\tau} - (\gamma_{\nu\rho}R\indices{^{\nu}_{\sigma\beta\tau}}\gamma^{\beta\sigma} + R\indices{^{\alpha}_{\rho\alpha\tau}})u^{\rho}u^{\tau} + 2(\bm{k}\cdot\bm{t})^{-1} \gamma_{\alpha\mu}u^{\mu}R\indices{^{\alpha}_{\rho\beta\tau}}u^{\rho}t^{\beta}u^{\tau}
\end{equation}
which brings us to the final form for the term $h_{\alpha}^{\beta}a\indices{^{\alpha}_{;\beta}}$
\begin{multline} \label{Eq:Term2}
 h_{\alpha}^{\beta}a\indices{^{\alpha}_{;\beta}}  = -\frac12h_{\alpha}^{\beta}(\gamma\indices{^{\alpha\sigma}_{;\beta}}\gamma_{\sigma\rho;\tau} + \gamma^{\alpha\sigma}\gamma_{\sigma\rho;\beta\tau})u^{\rho}u^{\tau} - \frac12 \gamma^{\alpha\sigma}\gamma_{\sigma\rho;\tau}(B\indices{^{\rho}_{\alpha}}u^{\tau} + B\indices{^{\tau}_{\alpha}}u^{\rho} )  +\frac12 (\gamma_{\nu\rho}R\indices{^{\nu}_{\sigma\beta\tau}}\gamma^{\beta\sigma} + R\indices{^{\alpha}_{\rho\alpha\tau}})u^{\rho}u^{\tau} \\ - (\bm{k}\cdot\bm{t})^{-1} \gamma_{\alpha\mu}u^{\mu}R\indices{^{\alpha}_{\rho\beta\tau}}u^{\rho}t^{\beta}u^{\tau} 
 - (\bm{k}\cdot\bm{t})^{-1}\gamma^{\alpha\sigma}\dot{\gamma}_{\sigma\rho}u^{\rho}\left[ (\gamma_{\lambda\mu}t^{\mu}u\indices{^{\lambda}_{;\alpha}}) + \dot{(\gamma_{\alpha\mu}t^{\mu})} \right]\\
   + (\bm{k}\cdot\bm{t})^{-2}(\dot{\gamma}_{\sigma\rho}u^{\sigma}u^{\rho}) \gamma_{\beta\nu}t^{\nu}[t^{\lambda} u\indices{^{\beta}_{;\lambda}} -u^{\lambda} t\indices{^{\beta}_{;\lambda}}] 
   + (\bm{k}\cdot\bm{t})^{-3} \dot{(\bm{k}\cdot\bm{t})}\dot{\gamma}_{\alpha\mu}u^{\alpha}[u^{\mu}+(\bm{k}\cdot\bm{t}) t^{\mu}].
\end{multline}

\subsection{$\dot{h}_{\alpha}^{\lambda}u\indices{^{\alpha}_{;\lambda}}$}
Our task here is somewhat simplified by the fact that we already calculated $\dot{h}_{\alpha}^{\lambda}$ in the last section.  We have
\begin{equation} \label{Eq:Term3}
\begin{aligned}
\dot{h}_{\alpha}^{\beta}u\indices{^{\alpha}_{;\beta}}
  & = \dot{(\bm{k}\cdot\bm{t})}(\bm{k}\cdot\bm{t})^{-3} [\gamma_{\alpha\mu}a^{\alpha}(2u^{\mu}+(\bm{k}\cdot\bm{t})t^{\mu}) + (\bm{k}\cdot\bm{t}) (\gamma_{\alpha\mu}u^{\mu}u\indices{^{\alpha}_{;\beta}}) t^{\beta}]
  - \frac12 (\bm{k}\cdot\bm{t})^{-2} \dot{\gamma}_{\alpha\mu}u^{\mu}[a^{\alpha} + (\bm{k}\cdot\bm{t})t^{\beta}u\indices{^{\alpha}_{;\beta}}] \\
    & \qquad + \frac12 (\bm{k}\cdot\bm{t})^{-2} \gamma^{\beta\sigma}\dot{\gamma}_{\sigma\rho}u^{\rho} [(\gamma_{\alpha\mu}u^{\mu}u\indices{^{\alpha}_{;\beta}}) + (\bm{k}\cdot\bm{t}) (\gamma_{\alpha\mu}t^{\mu}u\indices{^{\alpha}_{;\beta}})] - (\bm{k}\cdot\bm{t})^{-1} [\dot{t}^{\beta}(\gamma_{\alpha\mu}u^{\mu} u\indices{^{\alpha}_{;\beta}}) + a^{\alpha}\dot{(\gamma_{\alpha\mu}t^{\mu})}] \\
  & = \dot{(\bm{k}\cdot\bm{t})}(\bm{k}\cdot\bm{t})^{-3} [-\frac12 \dot{\gamma}_{\alpha\mu}u^{\alpha}(2u^{\mu}+(\bm{k}\cdot\bm{t})t^{\mu}) - \frac12 (\bm{k}\cdot\bm{t}) \dot{\gamma}_{\alpha\beta}u^{\alpha}t^{\beta}] 
  - \frac12 (\bm{k}\cdot\bm{t})^{-2} [-\frac12 \dot{\gamma}_{\alpha\mu}\gamma^{\alpha\sigma}\dot{\gamma}_{\sigma\nu}u^{\mu}u^{\nu} + (\bm{k}\cdot\bm{t}) \dot{\gamma}_{\alpha\mu}u^{\mu} t^{\beta}u\indices{^{\alpha}_{;\beta}}] \\
    & \qquad + \frac12 (\bm{k}\cdot\bm{t})^{-2} \gamma^{\beta\sigma}\dot{\gamma}_{\sigma\rho}u^{\rho}[-\frac12 \dot{\gamma}_{\mu\beta} u^{\mu} + (\bm{k}\cdot\bm{t})(\gamma_{\alpha\mu}t^{\mu}u\indices{^{\alpha}_{;\beta}})]
    -(\bm{k}\cdot\bm{t})^{-1}[-\frac12 \dot{\gamma}_{\alpha\beta}u^{\alpha}\dot{t}^{\beta} - \frac12 \gamma^{\alpha\sigma}\dot{\gamma}_{\sigma\nu}u^{\nu}\dot{(\gamma_{\alpha\mu}t^{\mu})}] \\
  & = -\dot{(\bm{k}\cdot\bm{t})}(\bm{k}\cdot\bm{t})^{-3} \dot{\gamma}_{\alpha\mu}u^{\alpha}[u^{\mu}+(\bm{k}\cdot\bm{t})t^{\mu}]
  + \frac12 (\bm{k}\cdot\bm{t})^{-1}\gamma^{\alpha\sigma}\dot{\gamma}_{\sigma\nu}u^{\nu}[\gamma_{\beta\mu}t^{\mu}u\indices{^{\beta}_{;\alpha}} + \dot{(\gamma_{\alpha\mu}t^{\mu})} + \gamma_{\alpha\beta}(u^{\lambda}t\indices{^{\beta}_{;\lambda}} - t^{\lambda}u\indices{^{\beta}_{;\lambda}})].
\end{aligned}
\end{equation}

\subsection{$h^{\mu}_{\rho}u\indices{^{\rho}_{;\tau}}(\delta^{\tau}_{\nu}-h^{\tau}_{\nu})(\delta^{\nu}_{\sigma}-h^{\nu}_{\sigma}) u\indices{^{\sigma}_{;\beta}} h^{\beta}_{\mu} = h^{\beta}_{\rho}u\indices{^{\rho}_{;\nu}}(\delta^{\nu}_{\sigma}-h^{\nu}_{\sigma})u\indices{^{\sigma}_{;\beta}}$}
Start with
\begin{equation}
 \begin{aligned}
   h^{\beta}_{\rho}u\indices{^{\rho}_{;\nu}}(\delta^{\nu}_{\sigma}-h^{\nu}_{\sigma})u\indices{^{\sigma}_{;\beta}} 
     & = (\bm{k}\cdot\bm{t})^{-2}h^{\beta}_{\rho}u\indices{^{\rho}_{;\nu}} \gamma_{\sigma\mu}\left[ u^{\mu}u^{\nu}+(\bm{k}\cdot\bm{t})(t^{\nu}u^{\mu} + t^{\mu}u^{\nu})  \right]u\indices{^{\sigma}_{;\beta}} \\
     & = (\bm{k}\cdot\bm{t})^{-2}h^{\beta}_{\rho} \left[ (\gamma_{\sigma\mu}u^{\mu}u\indices{^{\sigma}_{;\beta}})a^{\rho} + (\bm{k}\cdot\bm{t})\left( (\gamma_{\sigma\mu}u^{\mu}u\indices{^{\sigma}_{;\beta}})(t^{\nu}u\indices{^{\rho}_{;\nu}}) + (\gamma_{\sigma\mu}t^{\mu}u\indices{^{\sigma}_{;\beta}})a^{\rho} \right) \right] \\
     & = (\bm{k}\cdot\bm{t})^{-2}h^{\beta}_{\rho} \left[ \frac14 \dot{\gamma}_{\beta\sigma}\gamma^{\rho\lambda}\dot{\gamma}_{\lambda\mu} u^{\sigma}u^{\mu} - \frac12 (\bm{k}\cdot\bm{t})\left( \dot{\gamma}_{\beta\sigma}u^{\sigma}(t^{\nu}u\indices{^{\rho}_{;\nu}}) + \gamma^{\rho\nu}\dot{\gamma}_{\nu\lambda}u^{\lambda}(\gamma_{\sigma\mu}t^{\mu}u\indices{^{\sigma}_{;\beta}}) \right)  \right].
 \end{aligned}
\end{equation}
Adding and subtracting allows us to write this as
\begin{multline}
   h^{\beta}_{\rho}u\indices{^{\rho}_{;\nu}}(\delta^{\nu}_{\sigma}-h^{\nu}_{\sigma})u\indices{^{\sigma}_{;\beta}} 
     = \frac14 h^{\beta}_{\rho} \gamma^{\sigma\omega}\dot{\gamma}_{\beta\sigma}\gamma^{\rho\lambda}\dot{\gamma}_{\lambda\mu}(\delta^{\mu}_{\omega}-h^{\mu}_{\omega}) \\ 
     - \frac12 (\bm{k}\cdot\bm{t})^{-1} h^{\beta}_{\rho}\left[ \frac12\dot{\gamma}_{\beta\sigma}\gamma^{\rho\lambda}\dot{\gamma}_{\lambda\mu}(u^{\sigma}t^{\mu}+ u^{\mu}t^{\sigma})  +  \dot{\gamma}_{\beta\sigma}u^{\sigma}(t^{\nu}u\indices{^{\rho}_{;\nu}}) + \gamma^{\rho\nu} \dot{\gamma}_{\nu\mu} u^{\mu}(\gamma_{\sigma\lambda}t^{\lambda}u\indices{^{\sigma}_{;\beta}}) \right].
\end{multline}
Next expand the projection operator on the second line
\begin{equation}
 \begin{aligned}
   h^{\beta}_{\rho}u\indices{^{\rho}_{;\nu}}(\delta^{\nu}_{\sigma}-h^{\nu}_{\sigma})u\indices{^{\sigma}_{;\beta}} 
       & = \frac14 h^{\beta}_{\rho} \gamma^{\sigma\omega}\dot{\gamma}_{\beta\sigma}\gamma^{\rho\lambda}\dot{\gamma}_{\lambda\mu}(\delta^{\mu}_{\omega}-h^{\mu}_{\omega}) 
       - \frac12 (\bm{k}\cdot\bm{t})^{-1} \gamma^{\rho\lambda}\dot{\gamma}_{\rho\sigma}u^{\sigma}\left[  \dot{\gamma}_{\lambda\mu} t^{\mu} +\gamma_{\lambda\mu}t^{\nu}u\indices{^{\mu}_{;\nu}} + \gamma_{\nu\mu}t^{\nu}u\indices{^{\mu}_{;\lambda}} \right] \\
          & \qquad + \frac12 (\bm{k}\cdot\bm{t})^{-3}  \left[ \dot{\gamma}_{\beta\sigma} \dot{\gamma}_{\omega\mu}u^{\sigma}t^{\mu}\left(u^{\omega}u^{\beta}+(\bm{k}\cdot\bm{t})(u^{\omega} t^{\beta} + u^{\beta}t^{\omega})\right) \right. \\          
          & \qquad  + \dot{\gamma}_{\beta\sigma}u^{\sigma} (t^{\nu}u\indices{^{\rho}_{;\nu}}) \gamma_{\rho\omega} \left(u^{\omega}u^{\beta}+(\bm{k}\cdot\bm{t})(u^{\omega} t^{\beta} + u^{\beta}t^{\omega})\right)  \\
          & \qquad \left. +  \dot{\gamma}_{\omega\nu}u^{\nu}(\gamma_{\sigma\mu}t^{\mu}u\indices{^{\sigma}_{;\beta}}) \left(u^{\omega}u^{\beta}+(\bm{k}\cdot\bm{t})(u^{\omega} t^{\beta} + u^{\beta}t^{\omega})\right) \right] \\
       & = \frac14 h^{\beta}_{\rho} \gamma^{\sigma\omega}\dot{\gamma}_{\beta\sigma}\gamma^{\rho\lambda}\dot{\gamma}_{\lambda\mu}(\delta^{\mu}_{\omega}-h^{\mu}_{\omega}) 
       - \frac12 (\bm{k}\cdot\bm{t})^{-1} \gamma^{\rho\lambda}\dot{\gamma}_{\rho\sigma}u^{\sigma}\left[ \gamma_{\nu\mu}t^{\mu}u\indices{^{\nu}_{;\lambda}} + \dot{(\gamma_{\lambda\mu}t^{\mu})} + \gamma_{\lambda\mu} (t^{\nu}u\indices{^{\mu}_{;\nu}} - u^{\nu}t\indices{^{\mu}_{;\nu}}) \right] \\
          & \qquad + \frac12 (\bm{k}\cdot\bm{t})^{-3}  \left[ \dot{\gamma}_{\beta\sigma}u^{\sigma}(u^{\beta}+(\bm{k}\cdot\bm{t})t^{\beta}) (\dot{\gamma}_{\omega\mu}u^{\omega}t^{\mu}) + (\bm{k}\cdot\bm{t}) (\dot{\gamma}_{\beta\sigma}u^{\sigma}u^{\beta})(\dot{\gamma}_{\omega\mu}t^{\omega}t^{\mu}) \right. \\
          & \qquad  + \dot{\gamma}_{\beta\sigma}u^{\sigma}(u^{\beta}+(\bm{k}\cdot\bm{t})t^{\beta})  (\gamma_{\rho\omega}u^{\omega}u\indices{^{\rho}_{;\nu}}) t^{\nu} +  (\bm{k}\cdot\bm{t})(\dot{\gamma}_{\beta\sigma}u^{\sigma}u^{\beta})\gamma_{\rho\omega}t^{\omega} (t^{\nu} u\indices{^{\rho}_{;\nu}})  \\                    
          & \qquad \left. + \dot{\gamma}_{\omega\nu}u^{\nu}(u^{\omega}+ (\bm{k}\cdot\bm{t})t^{\omega})(\gamma_{\sigma\mu}a^{\sigma}t^{\mu}) + (\bm{k}\cdot\bm{t}) (\dot{\gamma}_{\omega\nu}u^{\nu}u^{\omega})\gamma_{\sigma\mu}t^{\mu}(t^{\beta}u\indices{^{\sigma}_{;\beta}}) \right] \\
       & = \frac14 h^{\beta}_{\rho} \gamma^{\sigma\omega}\dot{\gamma}_{\beta\sigma}\gamma^{\rho\lambda}\dot{\gamma}_{\lambda\mu}(\delta^{\mu}_{\omega}-h^{\mu}_{\omega}) 
       - \frac12 (\bm{k}\cdot\bm{t})^{-1} \gamma^{\rho\lambda}\dot{\gamma}_{\rho\sigma}u^{\sigma}\left[ \gamma_{\nu\mu}t^{\mu}u\indices{^{\nu}_{;\lambda}} + \dot{(\gamma_{\lambda\mu}t^{\mu})} + \gamma_{\lambda\mu} (t^{\nu}u\indices{^{\mu}_{;\nu}} - u^{\nu}t\indices{^{\mu}_{;\nu}}) \right] \\
          & \qquad + \frac12 (\bm{k}\cdot\bm{t})^{-3}  \left[ \dot{\gamma}_{\beta\sigma}u^{\sigma}(u^{\beta}+(\bm{k}\cdot\bm{t})t^{\beta}) (\dot{\gamma}_{\omega\mu}u^{\omega}t^{\mu}) + (\bm{k}\cdot\bm{t}) (\dot{\gamma}_{\beta\sigma}u^{\sigma}u^{\beta})(-2\gamma_{\omega\mu}t^{\omega}\dot{t}^{\mu}) \phantom{\frac12} \right. \\
          & \qquad  + \dot{\gamma}_{\beta\sigma}u^{\sigma}(u^{\beta}+(\bm{k}\cdot\bm{t})t^{\beta})  \left(-\frac12 \dot{\gamma}_{\nu\omega}u^{\omega}\right) t^{\nu} +  (\bm{k}\cdot\bm{t})(\dot{\gamma}_{\beta\sigma}u^{\sigma}u^{\beta})\gamma_{\rho\omega}t^{\omega} (t^{\nu} u\indices{^{\rho}_{;\nu}})  \\                    
          & \qquad \left. + \dot{\gamma}_{\omega\nu}u^{\nu}(u^{\omega}+ (\bm{k}\cdot\bm{t})t^{\omega}) \left(-\frac12 \dot{\gamma}_{\sigma\mu}u^{\sigma}t^{\mu}\right) + (\bm{k}\cdot\bm{t}) (\dot{\gamma}_{\omega\nu}u^{\nu}u^{\omega})\gamma_{\sigma\mu}t^{\mu}(t^{\beta}u\indices{^{\sigma}_{;\beta}}) \right] \\    
       & = \frac14 h^{\beta}_{\rho} \gamma^{\sigma\omega}\dot{\gamma}_{\beta\sigma}\gamma^{\rho\lambda}\dot{\gamma}_{\lambda\mu}(\delta^{\mu}_{\omega}-h^{\mu}_{\omega}) 
       - \frac12 (\bm{k}\cdot\bm{t})^{-1} \gamma^{\rho\lambda}\dot{\gamma}_{\rho\sigma}u^{\sigma}\left[ \gamma_{\nu\mu}t^{\mu}u\indices{^{\nu}_{;\lambda}} + \dot{(\gamma_{\lambda\mu}t^{\mu})} + \gamma_{\lambda\mu} (t^{\nu}u\indices{^{\mu}_{;\nu}} - u^{\nu}t\indices{^{\mu}_{;\nu}}) \right] \\
          & \qquad + (\bm{k}\cdot\bm{t})^{-2}  (\dot{\gamma}_{\beta\sigma}u^{\sigma}u^{\beta}) \gamma_{\omega\mu}t^{\omega} [t^{\nu}u\indices{^{\mu}_{;\nu}}- u^{\nu}t\indices{^{\mu}_{;\nu}}],
 \end{aligned}
\end{equation}
where we have made use of the fact that $\gamma_{\mu\nu}t^{\mu}t^{\nu}$ is constant to write
\begin{equation}
 \dot{\gamma}_{\sigma\mu}t^{\sigma}t^{\mu} = -2\gamma_{\sigma\mu}t^{\sigma}\dot{t}^{\mu}.
\end{equation}
We are left with 
\begin{multline} \label{Eq:Term4}
 h^{\beta}_{\rho}u\indices{^{\rho}_{;\nu}}(\delta^{\nu}_{\sigma}-h^{\nu}_{\sigma})u\indices{^{\sigma}_{;\beta}}   
         = \frac14 h^{\beta}_{\rho} \gamma^{\sigma\omega}\dot{\gamma}_{\beta\sigma}\gamma^{\rho\lambda}\dot{\gamma}_{\lambda\mu}(\delta^{\mu}_{\omega}-h^{\mu}_{\omega}) \\
                - \frac12 (\bm{k}\cdot\bm{t})^{-1} \gamma^{\rho\lambda}\dot{\gamma}_{\rho\sigma}u^{\sigma}\left[ \gamma_{\nu\mu}t^{\mu}u\indices{^{\nu}_{;\lambda}} + \dot{(\gamma_{\lambda\mu}t^{\mu})} + \gamma_{\lambda\mu} (t^{\nu}u\indices{^{\mu}_{;\nu}} - u^{\nu}t\indices{^{\mu}_{;\nu}}) \right] 
                    + (\bm{k}\cdot\bm{t})^{-2}  (\dot{\gamma}_{\beta\sigma}u^{\sigma}u^{\beta}) \gamma_{\omega\mu}t^{\omega} [t^{\nu}u\indices{^{\mu}_{;\nu}}- u^{\nu}t\indices{^{\mu}_{;\nu}}].
\end{multline}
With Eqs. (\ref{Eq:Term1}), (\ref{Eq:Term2}), (\ref{Eq:Term3}), and (\ref{Eq:Term4}) and some final rearranging, we finally have the sought-after solution
\begin{multline}
 \dot{\Theta} = -B\indices{^{\mu}_{\nu}}B\indices{^{\nu}_{\mu}} - \frac12 R\indices{_{\beta\tau\alpha}^{\nu}} (g^{\alpha\beta}g_{\nu\lambda}+\gamma^{\alpha\beta}\gamma_{\nu\lambda})u^{\tau}u^{\lambda}
 -\frac12 \gamma^{\alpha\sigma}\gamma_{\sigma\rho;\tau} (B\indices{^{\rho}_{\alpha}}u^{\tau}+B\indices{^{\tau}_{\alpha}}u^{\rho}) \\
 -\frac12 h^{\beta}_{\alpha}\left[\gamma^{\alpha\sigma}\gamma_{\sigma\rho;\beta\tau}u^{\rho}u^{\tau} + \gamma\indices{^{\alpha\sigma}_{;\beta}}\dot{\gamma}_{\sigma\rho}u^{\rho} + \frac12 \gamma^{\sigma\omega}\dot{\gamma}_{\beta\sigma}\gamma^{\alpha\lambda}\dot{\gamma}_{\lambda\mu}(\delta^{\mu}_{\omega}-h^{\mu}_{\omega}) \right].
\end{multline}



%

\end{document}